\documentclass[sigconf]{acmart}

\AtBeginDocument{%
  \providecommand\BibTeX{{%
    \normalfont B\kern-0.5em{\scshape i\kern-0.25em b}\kern-0.8em\TeX}}}

\setcopyright{acmcopyright}
\copyrightyear{2018}
\acmYear{2018}
\acmDOI{XXXXXXX.XXXXXXX}

\acmConference[Conference acronym 'XX]{Make sure to enter the correct
  conference title from your rights confirmation email}{June 03--05,
  2018}{Woodstock, NY}
\acmBooktitle{Woodstock '18: ACM Symposium on Neural Gaze Detection,
 June 03--05, 2018, Woodstock, NY} 
\acmPrice{15.00}
\acmISBN{978-1-4503-XXXX-X/18/06}

\acmSubmissionID{6484}

\begin{document}

\title[Understanding Generative AI in Art]{Understanding Generative AI in Art: An Interview Study with Artists on G-AI from an HCI Perspective}

\author{Jingyu Shi}
\orcid{0000-0001-5159-2235}
\email{shi537@purdue.edu}
\affiliation{%
  \institution{Purdue University}
  \city{West Lafayette}
  \state{Indiana}
  \country{USA}
}

\author{Rahul Jain}
\email{jain348@purdue.edu}
\orcid{0009-0001-3723-5482}
\affiliation{%
  \institution{Purdue University}
  \city{West Lafayette}
  \state{Indiana}
  \country{USA}
}

\author{Runlin Duan}
\orcid{0000-0001-8256-6419}
\email{duan95@purdue.edu}
\affiliation{%
  \institution{Purdue University}
  \city{West Lafayette}
  \state{Indiana}
  \country{USA}
}

\author{Karthik Ramani}
\orcid{0000-0001-8639-5135}
\email{ramani@purdue.edu}
\affiliation{%
  \institution{Purdue University}
  \city{West Lafayette}
  \state{Indiana}
  \country{USA}
}

\renewcommand{\shortauthors}{Trovato and Tobin, et al.}

\begin{abstract}
  The emergence of Generative Artificial Intelligence (G-AI) has changed the landscape of creative arts with its power to compose novel artwork and thus brought ethical concerns.
  Despite the efforts by prior works to address these concerns from technical and societal perspectives, there exists little discussion on this topic from an HCI point of view, considering the artists as human factors.
  We sought to investigate the impact of G-AI on artists, understanding the relationship between artists and G-AI, in order to motivate the underlying HCI research.
  We conducted semi-structured interviews with artists ($N=25$) from diverse artistic disciplines involved with G-AI in their artistic creation.
  We found (1) a dilemma among the artists, (2) a disparity in the understanding of G-AI between the artists and the AI developers(3) a tendency to oppose G-AI among the artists.
  We discuss the future opportunities of HCI research to tackle the problems identified from the interviews.
\end{abstract}




\received{20 February 2007}
\received[revised]{12 March 2009}
\received[accepted]{5 June 2009}

\maketitle
\section{Introduction}
The emergence of Generative Artificial Intelligence (henceforth, G-AI) has changed the landscape of the realm of creative arts~\cite{newton2023ai}, encompassing fields like literature~\cite{li2019storygan, peng2019poetry}, music~\cite{li2021inco,zhu2022quantized}, painting~\cite{karras2019style,brock2018large}, among others.
G-AI has democratized the creative process by empowering individuals, regardless of their artistic proficiency, to engage in diverse forms of creative art without the need for extensive training or skill acquisition, as was previously required.
For instance, individuals can now effortlessly transfer the styles of recognized painters to their own artwork~\cite{karras2019style}, generate an artistic picture with simply a piece of textual description~\cite{ramesh2022hierarchical}, or interactively write fiction with a chat-bot~\cite{brown2020language}.
Such accessibility has opened up novel avenues for designing, creating, and refining artwork across various domains.

However, this proliferation of generative power in artwork has unwittingly opened Pandora's box, as G-AI rapidly surpassed human capabilities~\cite{elkins2020can,haase2023artificial,ward2017AI} and raised ethical dilemmas.
Particularly, oppositions criticize the usage of G-AI for its invasion of originality, creativity, and copyrights of human art creators~\cite{DIEN2023108621, noci2023merging, doi/10.2759/570559, liu2023arguments}.
This is no novel topic for the AI community, where debates on the ethics of AI have endured for decades.
Nevertheless, the ascendancy of G-AI is factually more powerful than any AI system before, and therefore breaches a line beyond societal acceptance, resulting in escalated and unprecedented discussions~\cite{roose2022ai,waters2023generative}.

Under this context, prior works put efforts into locating, justifying, or questioning this breached line from diverse perspectives~\cite{bran2023emerging}.
To leverage the negative impact of G-AI, many have proposed, viewing from the technical side, e.g. reducing the use of human-original content as training data for AIs~\cite{machineunlearning,abrahamsen2023inventing}.
Alternatively, some elucidate the mechanism of G-AI and consequently conclude practical boundaries between plagiarism and legitimate reproduction~\cite{sarkar2023exploring,todorov2019game}.
Besides the technical discussions, other researchers approach from the societal aspect, revealing the exploitation of the artists~\cite{ghosh2022can}, which is not only posed by G-AI itself but also induced by its owners or users.
The community realized that the deployment of G-AI is far from a mature and harmonious symbiosis between artists and G-AI.
More efforts are expected to shift G-AI technically, societally, and philosophically~\cite{boden2009generative} towards an accommodating integration~\cite{AIArtists}.

Despite the technical and societal progression, little prior work approaches from the artists' perspective.
We are shown how human factors are still indispensable~\cite{stackpole2023why, markuckas2022question} and preferred~\cite{bellaiche2023humans, 10.1145/3334480.3382892} in the context of art.
To motivate the community towards a Human-GAI symbiosis, we directed our research towards a critical aspect overlooked in the discourse: the artists, who constitute a crucial human factor and potential victims of this revolution.
We aim to derive insights on future research opportunities and challenges regarding the interactions between artists and GAI, by investigating the subjective answers to the following questions:

\begin{itemize}
  \item \textbf{How are artists' careers changed by G-AI?}
  \item \textbf{What are artists' perspectives on G-AI?}
  \item \textbf{How do we bridge artists' perspectives with future research on G-AI?}
\end{itemize}

To this end, we undertake a series of in-depth interviews with artists representing various disciplines.
We conduct qualitative analyses of these interviews, which elicit the following key insights:
\begin{enumerate}
    \item Artists (or art creators) are forced into an N-Person Prisoner's Dilemma~\cite{hamburger1973n}, where they feel \textbf{compelled to use G-AI} to enhance efficiency, refine ideation, adapt styles, etc., \textbf{in order to survive in the competition amidst their peers}. 
This invasive deployment of G-AI leads to its widespread adoption among artists.
Under this circumstance, \textbf{all surviving artists resort to the use of G-AI}, with or without awareness of the potential leak of their work, regardless of the user-friendliness of the GAI systems.
We identify this compromising usage of GAI as a negative factor blocking possible improvement in the integration of GAI.

\item The Rashomon Effect~\cite{heider1988rashomon} under the topic of G-AI in art remains unaddressed.
Despite the efforts put into elucidating the nature of the G-AI algorithms and dispelling concerns of intrusiveness and understanding of art akin to humans, \textbf{artists are unable and unwilling to comprehend the technical perspective}, regardless of how straightforward the underlying math and equations are presented.
On the other side of the coin, \textbf{current AI research overlooks the artists' perspectives} and how they should handle the integration with G-AI without any technical knowledge of the mechanism of G-AI.
Deep research is anticipated to take both perspectives into consideration and bring about the symbiosis between G-AI and artists.
\item A tendency to oppose G-AI has grown among artists.
Such opposition is to some extent misled.
While a few comprehend the notion of profit shifting~\cite{ghosh2022can} and how their earnings are appropriated by the commercial owners and developers of G-AI, \textbf{many direct their blame toward the technology itself}.
This aversion towards the technology, rather than its utilization, may bring both sides to a negative confrontation and will eventually result in a lose-lose situation.
\end{enumerate} 

Inspired by the findings, we further discuss future research on Human-G-AI symbiosis.
We conclude with opportunities and challenges from both G-AI and artists' perspectives around (1) attribution of credit in creativity, originality, and copyright, (2) rules, regulations, and consensus of using G-AI among the artist community, (3) novice-friendly education toward G-AI for artists to understand G-AI mechanism, and (4) technical methodologies for prior-plagiarism and post-plagiarism solutions.
\section{Background}
In this section, we will first go through the technical development of G-AI and shed light on how technical development has changed the game and atmosphere in modern creative art.
Subsequently, we will reveal the impact of G-AI on the art community, starting from the ethical concerns addressed by society and academia.
We then traverse through prior works focusing on Human-G-AI symbiosis, approaching either the technical solutions or societal concerns.
Finally, we conclude the motivations and perspectives of our interviews from the overall landscape of G-AI.
\subsection{Development in Generative AI for Art}
Generative Art is by no means a novel concept~\cite{boden2009generative}.
It thrives on the concept of algorithmic creativity, wherein 
\textbf{artists collaborate with computer algorithms to create artistic work}.
Owing to its rapid development in the last decade, G-AI has become the mainstream tool for generative art.
Our focus here is confined to the recent surge in G-AI advancements, which have resulted in transformative changes in the field.

The introduction of Generative Adversarial Networks (GANs)~\cite{goodfellow2014generative} has had a profound impact on the field of generative AI and thus the field of generative art.
It enables users to create images based on the training data they input.
With efforts in stabilized~\cite{radford2015unsupervised, salimans2016improved} and enhanced~\cite{karras2017progressive} quality of GANs, the successors of GAN open a broader vision of applications of G-AI in generative art.
Conditional GANs~\cite{mirza2014conditional}, extend the GAN framework with conditional embedding, \textbf{allowing the generation of specific output based on conditioning information}, leading to foundational applications like image-to-image~\cite{Zhu_2017_ICCV}, text-to-image generation~\cite{Liao_2022_CVPR}, audio-to-image~\cite{wan2019towards} translation, etc.
Based on the foundational applications, later works enable more customizable and realistic outputs for specific tasks such as style transfer and fusion~\cite{karras2019style}, super-resolution image generation~\cite{brock2018large}, video generation~\cite{wang2020imaginator}, etc.
Applications have been introduced into the realm of art, such as photography~\cite{kamran2021rv}, painting~\cite{gao2020painting}, graphic design~\cite{li2020attribute}, music~\cite{li2021inco, kulkarni2019survey, zhu2022quantized}, etc.

Recent development in Natural Language Processing (NLP), particularly the research in Large Language Models (LLMs), has changed the game in literature arts~\cite{yenduri2023generative}.
Models such as Generative Pre-trained Transformer (GPT) and its successors~\cite{radford2018improving, radford2019language, keskar2019ctrl, brown2020language}, and T5~\cite{raffel2020exploring}, and BERT~\cite{devlin2018bert} \textbf{prowess in a wide array of language tasks} such as text generation, writing assistance, translation, dialogue development, narrative design, etc.
The utilization of large models (also known as foundation models) soon \textbf{extended into the vision community}.
OpenAI~\cite{openaiwebsite} bridged between image embedding with LLMs and released Contrastive Language-Image Pre-Training (CLIP)~\cite{radford2021learning}, which can perform visual analogies and knowledge transferring among modalities.
CLIP opens a bi-directional avenue between text and image and thus empowers the creation of literature and visual arts.
Subsequently, follow-up research based on foundation models including Imagen~\cite{saharia2022photorealistic}, DALL.E~\cite{ramesh2021zeroshot, ramesh2022hierarchical}, and Midjourney~\cite{midjourneywebsite}, further showcases the continued abilities of G-AI in the realm of artwork.
The powerful applications of large models opened up by large companies created \textbf{user-friendly, easily accessible, and scalable tools for assistance, collaboration, and end-to-end creation of most forms of artwork}. Besides, in the wake of the concept of Diffusion Model~\cite{sohldickstein2015deep}, Diffusion-based content generation~\cite{ho2020denoising, Rombach_2022_CVPR, NEURIPS2022_ec795aea, nichol2022glide} came to the topic.
Diffusion models grant the users generated artwork with content superior to their predecessors~\cite{NEURIPS2021_49ad23d1}.
By such development in G-AI, it is concluded that \textbf{artwork is revolutionarily democratized}~\cite{newton2023ai}.

\subsection{AI Ethics in Arts}

\begin{figure*}[htp]
    \centering
    \includegraphics[width=.9\textwidth]{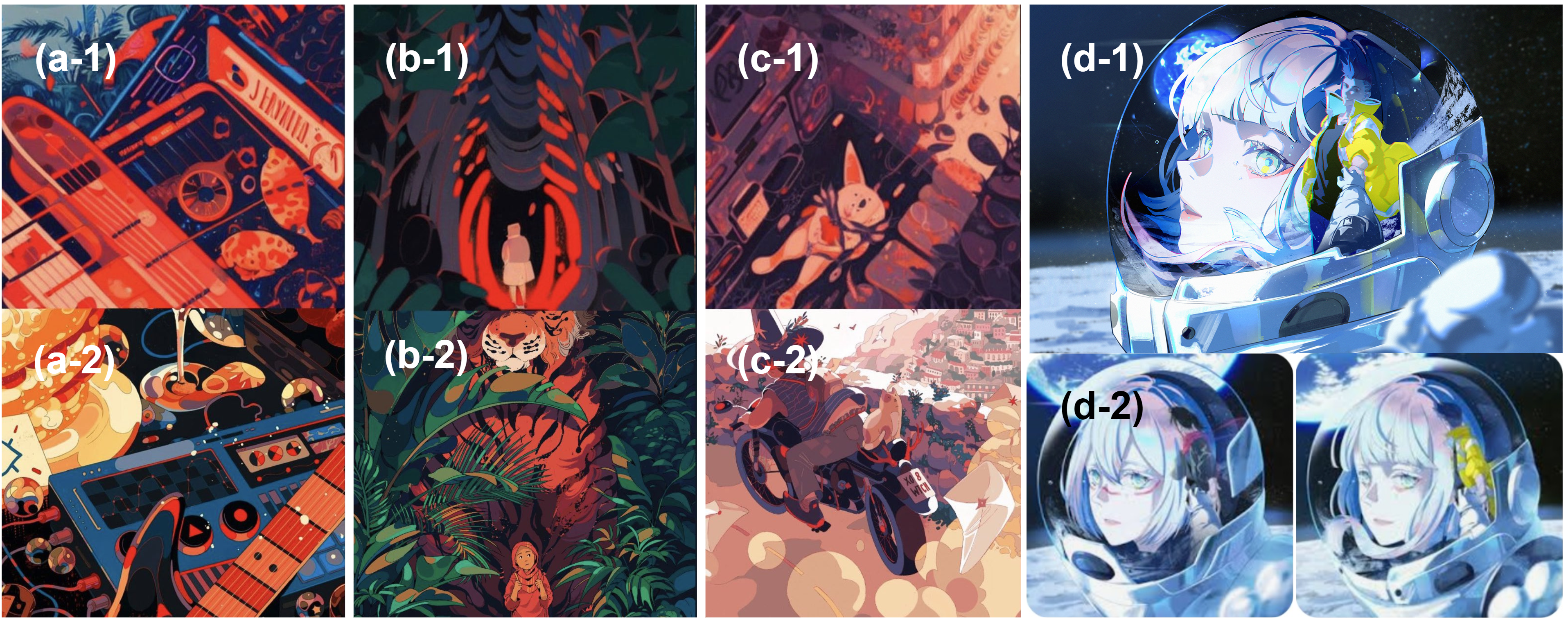}
    \caption{Examples of G-AI Plagiarisms of Illustrations from Human Artists. (a,b,c,d-1) are from human artists, and (a,b,c,d-2) are from G-AI generation. Obvious style and content transferring can be identified from (a,b,c)~\cite{debleewebsite, debleepost}. Identical content and style plagiarising can be observed from (d)~\cite{liflagwebsite, liflagpost}. All pictures and comparisons are from the artists' social media posts.}
    \label{fig:plagiarism}
\end{figure*}

The realm of AI ethics~\cite{liao2020ethics} has been a subject of prolonged and extensive discussion, covering the fields of AI fairness and bias~\cite{selbst2019fairness, mehrabi2021survey, NIPS2016_9d268236}, AI trustworthiness~\cite{brundage2020toward, brundage2018malicious}, and AI privacy~\cite{al2019privacy, danezis2015privacy}, among others.
Within this domain, the impact on art by G-AI has brought new and heated discussion from diverse perspectives.

Given G-AI's inherent generative capabilities, its creativity has been questioned and even opposed by part of the artist community.
Oppositions are voiced in the forms of manifestos against the widespread of G-AI for art~\cite{manifesto}, public inquisitions~\cite{10.1145/3592981} into or accusations~\cite{chayka2023is} of potential plagiarism by G-AI to the artists, and heated debates over the essence of creativity by G-AI~\cite{Baio2023Invasive}.
As a result, research has focused on this phenomenon, seeking to find the root of this controversy.
In this regard, we identify and elaborate on two major concerns that motivate prior works: G-AI plagiarism and G-AI's impact on the art market.

\subsubsection{G-AI Plagiarism}
The concern of plagiarism by or via G-AI appeared not only in creative art but also in the fields of education~\cite{francke2019potential} or academia~\cite{DIEN2023108621}.
Researchers have proposed various approaches to defining plagiarism by G-AI, drawing from notions of intellectual property~\cite{noci2023merging}, copyrights~\cite{doi/10.2759/570559}, and the very essence of creativity~\cite{liu2023arguments, sarkar2023exploring}.

A prevailing belief posits that G-AI does not truly learn or comprehend art in the way humans do, but rather operates within high-dimensional spaces by manipulating parameters~\cite{wu2023art, zylinska2020ai_not_creation, 10.1145/3475799}.
As an illustrative analogy, G-AI is likened to Candy Crush, wherein it engages in projections, combinations, and permutations of high-dimensional vectors~\cite{zylinska2020ai_not_creation}.
Consequently, some argue that G-AI's capacity to summarize and reorganize pre-existing materials results in a denial of genuine artistic creation or innovation~\cite{Baio2023Invasive}.
It is also shown by some work that current G-AI cannot replace human creativity and engagement in artwork~\cite{bellaiche2023humans}.
According to this perspective, G-AI-generated artworks cannot be considered plagiarized if they or their users do not intentionally replicate existing works to appear novel.

The opposite voice has it that based on established aesthetic criteria~\cite{mikalonyte2023art} or the crowd concept of art~\cite{mikalonytefolk}, G-AI can in principle create arts~\cite{mikalonyte2022can, coeckelbergh2023work}
It is argued by~\cite{liu2023arguments} that the definition of creativity in AI does not have to be dominated by humans and that AI has its unique definition of originality.
The possibility of G-AI generating novel artistic pieces raises concerns that artists may face challenges in asserting their originality.
Derived from research in literary criticism, the history of art, and copyright law, recent arguments by Advait Sarkar~\cite{sarkar2023exploring} emphasize that creativity and originality should be regarded as attributes of a process, an author, or a viewer rather than mere notations.
We conclude from above that the ongoing debates on the definition of AI plagiarism can be essentially traced back to the diverse perspectives that shape the definition of creativity in the context of G-AI's impact on art.

\subsubsection{G-AI's impact on the art market}
The changes in the art market brought by G-AI have resulted in fear that G-AI may gradually take away artists' markets and jobs.
Recent studies ~\cite{ali2023constructing} show that users of G-AI identify the risk of job replacement in general.
With AI-generated content (AIGC) showing up in the art market and gaining increasing recognition and acceptance~\cite{haase2023artificial, gabe2018up}, some believe that jobs involving a substantial element of design or writing are at risk~\cite{waters2023generative, zarifhonarvar2023economics}.
While negative voices towards G-AI taking away the artists' market are heard, some researchers~\cite{doi:10.1126/science.adh4451, zarifhonarvar2023economics} see the threat of job replacement in short terms but the enabling new forms of creative labor and reconfiguration of the art economy as the ultimate result.

Researches also shed light on how G-AI is been used as a tool to transfer the profit from the artists to the owners of G-AI~\cite{ghosh2022can}.
Put on a larger scope, the concern of the transferred profit may be taken advantage of by incumbents that control key inputs or adjacent markets towards unfair methods of competition~\cite{FTC2023generative}, the concentration of power~\cite{bommasani2022opportunities}, where all resources are controlled by the owners of G-AI.
It is also pointed out that the missing regulation in the market of GAI is also worsening the threats to the art market~\cite{wach2023dark}.
It remains uncertain how to tackle the revolution in the art market brought by G-AI in a nuanced approach that recognizes the reaction of artists and enables the capability of G-AI at the maximum in this domain.

\subsubsection{Related Prior Works}
Gaining insights from the ongoing debates, researchers established outlooks on the future progression addressing the concerns aforementioned~\cite{10.1145/3475799}.
Efforts have been made from the technical side, tweaking the structures, components, or data of AI to mitigate its threats to the art community.
In order to prevent possible breaches of intellectual property, one can train the G-AI models without the use of artists' data~\cite{abrahamsen2023inventing}.
Solutions are also given to address AI plagiarism by classifying instances of infringement~\cite{elgammal2018picasso, 10.1145/3503161.3548433, frye2022should, abd2019artificial}  and remedy upon detection by removing the learned content~\cite{machineunlearning}.
Moreover, to foster acceptance of G-AI among artists, many advocates to raise awareness of the G-AI mechanism, process, and usage~\cite{hutson2023generative, hutson2023content, 10.1145/3478432.3499157, zylinska2020ai_teach, hertzmann2022toward} among the crowd of artists.

Besides, the research has recognized the necessity of studying the social impact of G-AI in the context of arts~\cite{li2020research, franceschelli2023creativity, ghosh2022can}.
Recent works delve into the negative societal implications of G-AI~\cite{bird2023typology, flick2022ethics, ghosh2022can}and subsequently address the integrity of G-AI product~\cite{tomlinson2023chatgpt, abd2019artificial}, suggest practices of using G-AI~\cite{hutson2023content, hedges2023artificial}, and propose dedicated laws and regulations~\cite{zhang2023research, 10.1145/3593013.3594067}.

Despite the endeavors to tweak and regulate G-AI for the art society or educate the artists on what G-AI is and how to use it, limited research examines the perspective of artists themselves.
Evidence has demonstrated G-AI's poor performance in attaining the creation of art on its own~\cite{ajani2022human, markuckas2022question} and reaffirmed the irreplaceable~\cite{stackpole2023why, markuckas2022question} and preferred~\cite{bellaiche2023humans, 10.1145/3334480.3382892} human factors in artworks.
Human factors are still indispensable for art, at least at the current stage of G-AI development.

To shed light on the emergent landscape of G-AI with human factors in art, Jane Adams~\cite{adams2022one} shares a unique piece of reflection on the matter, as both an artist and a computer scientist.
Research by Patti Pente~\cite{pente2022artificial} also discusses the emergence of G-AI and art from the views of two artists.
Both pieces of research highlight that this human-GAI encounter challenges both artists and G-AI researchers to reach a level of emergence or symbiosis where the creativity and power of humans and G-AI both get to remain.
We propose to pose this symbiosis of G-AI and artists as an HCI problem and identify a missing piece in current research on the interactions and relationships between artists and G-AI.
To this end, we conduct an interview with artists across diverse disciplines to gain insights into existing challenges in Human-GAI interactions and to develop a comprehensive understanding of the role of human factors in the envisioned symbiosis.


\section{Methodology}
\subsection{Pilot Study}
To ground our research in the interviews, we conducted a pilot interview study with four artists ($pilot_1$ to $pilot_4$), two specializing in comics, one in music production, and one in essay writing, respectively.
The pilot interviews are conducted in the form of free talks on the topic of G-AI.
Each interviewee is asked to talk freely about anything related to G-AI they have in mind.
This pilot interview aims to examine the fundamental aspects of artists' understanding of G-AI, explore concerning topics in the context of artist-G-AI interactions, and provide insights into designing a series of in-depth interview questions.

All pilot users report that there have been major changes in their creation process due to the burst of G-AI development, addressing the need for a thorough analysis of how the emergence of G-AI plays a role in the changes.
We also see differences in user experience among the domains from the pilot answers, specifically, their distinctive usage of G-AI such as ideation assistance ($pilot_3$), draft generation ($pilot_1$ and $pilot_2$), variation application ($pilot_4$), among others.
The usage is a critical part of interactions between G-AI and artists as its users, we target the user experience of artists as one aspect of our interview.
Moreover, we identify a strong tendency of the pilot users to express their subjective opinions on G-AI.
All pilot users express their attitudes towards the current deployment of G-AI in art, either positive:
\begin{quote}
    \textit{"In one of my recent projects, I used a generative AI model to draw the fur of a red panda. The texture it generated was so good. I would have drawn this way if I had known how."} - $pilot_1$
\end{quote}

or negative:
\begin{quote}
    \textit{"Some collectors and galleries seem to be more intrigued by the novelty of AI-generated pieces. They think AI can replace painters and overlook the efforts and skill[s] [required for classical art]. This has made it increasingly difficult for me and others to gain recognition and sustain our careers."} - $pilot_2$
\end{quote}

Pilot users also show a diverse extent of expectations of future use of G-AI, either optimistic:
\begin{quote}
    \textit{"It is already powerful enough. I can't even imagine it being stronger. I am looking forward to seeing what more it can help me with."} - $pilot_1$
\end{quote}

permissive:
\begin{quote}
    \textit{"Well, if it keeps learning and stealing our work and style, it will get bigger. That might kill us [metaphorically in the market] though."}- $pilot_2$
\end{quote}

or observational:
\begin{quote}
    \textit{"I don't know, man. I don't really think it is that helpful now. It might change soon."} - $pilot_4$
\end{quote}.

To this end, we see the significance of artists' subjective opinions on and expectations of G-AI in this research.

Based on the responses, we refined our interview questions.
We focus on conversations with the artists to elaborate on their thoughts regarding how their artistic creation process and thus their career have been changed by G-AI, what their opinions on and expectations of G-AI are, and what they would like the G-AI developers to acknowledge.
Meanwhile, we designated follow-up questions that can elaborate on the previous questions to dig deeper into individual contexts.
We end up with four sections of questions: \textbf{experience, understanding, opinion, and expectation}.
We make sure the story flow is consistent in each interview by arranging the questions in an incremental and deepening tune.

\subsection{Participants}
We recruited artists who claim to have used G-AI or been affected by G-AI across multiple domains including visual arts, studio arts, and literature.
Specifically, our initial recruitment targets include university students from related departments, freelance artists from the online job markets, artists from well-recognized online communities, and referrals from other participants.
We eventually found 25 (P1-P25) participants from five countries agreeing to the interview.
The demographics of the participants are listed in ~\autoref{tab:participants}
\begin{table*}[htp]
  \caption{Demographics of the Participants}
  \label{tab:participants}
  \begin{tabular}{cccl}
    \toprule
    No. & Occupation & Country & Experience with G-AI\\
    \midrule
1	&	mangaka	&	Japan	&	generating manga	\\
2	&	mangaka	&	Japan	&	generating manga	\\
3	&	mangaka	&	Japan	&	modifying manga	\\
4	&	mangaka	&	China	&	ideation from generated manga character design	\\
5	&	mangaka	&	China	&	ideation from generated manga character design	\\
6	&	music producer	&	China	&	authoring beats	\\
7	&	music producer	&	China	&	generating remixes	\\
8	&	photographer	&	US	&	generating photographs	\\
9	&	photographer	&	China	&	generating photographs	\\
10	&	photographer	&	China	&	modifying photographs	\\
11	&	poster designer	&	China	&	generating post designs	\\
12	&	poster designer	&	US	&	generating post designs	\\
13	&	contemporary painter	&	China	&	generating digital paintings from text	\\
14	&	contemporary painter	&	China	&	generating digital paintings from text \\
15	&	contemporary painter	&	China	&	modifying painting	\\
16	&	contemporary painter	&	Korea	&	generating for ideation	\\
17	&	fiction writer	&	China	&	ideation with science knowledge	\\
18	&	news writer	&	US	&	modifying the news	\\
19	&	street artist	&	Canada	&	ideation from generated painting	\\
20	&	street artist	&	China	&	ideation from generated painting	\\
21	&	digital designer	&	China	&	generating web design for shopping websites	\\
22	&	digital designer	&	China	&	generating web design for shopping websites	\\
23	&	digital designer	&	China	&	ideation for advertisement design	\\
24	&	screenwriter	&	US	&	protesting against G-AI for screenwriting	\\
25	&	screenwriter	&	US	&	protesting against G-AI for screenwriting\\
  \bottomrule
\end{tabular}
\end{table*}

\subsection{Data Collection and Analysis}
We conducted our semi-structured interviews with the participants on Zoom, with each interview done individually in approximately 50 minutes.
Prior to the formulated questions from the pilot study, we asked the participants about their background in art, their high-level understanding of their specialization in art, and their broader experience through their careers in art.
After the formulated questions, we motivated the participants to give a free talk about G-AI as we did in the pilot study.

Interviews were recorded and transcribed automatically and then corrected manually.
We analyze the data following the constant comparative method~\cite{glaser1968discovery} with triangulation~\cite{triangulation2014use}.
three researchers who have partly designed and conducted the interviews went through the following analysis process, identically but separately.
First, they familiarized themselves by reviewing the recordings and taking necessary notes regarding the responses.
Then, they utilized NVivo, a qualitative evaluation software to assign codes to three different subsets of the transcripts.
Next, with more other interviews iterated, they continuously compared and refined the merging codes of the later transcripts.
Finally, three researchers merged their work by triangulating from their sets of codes, refining through discussion, and reapplying the codes to the transcripts.
We present our findings in the following section.
\section{Results}
In this section, we illustrate our results and findings from the interviews.
We aim to understand the aforementioned questions: (1) How are artists’ careers changed by G-AI subjectively? (2) What are artists’ perspectives on G-AI? and (3) How do we bridge artists’ perspectives with future research on G-AI?
With the questions in mind, we identify, from the interview analysis, an N-Person Prisoner's Dilemma where the artists compromise their general user experience with G-AI to compete with each other, a Rashomon in the observation where the opinions and understanding of G-AI vary across stances, and a misled resistance from the artists towards G-AI itself rather than the misusage of G-AI, posing potential harm towards the future development of G-AI.

\subsection{An N-Person Prisoner's Dilemma}

\subsubsection{Efficiency, variation, and ideation from G-AI}
Participants had positive feedback on the capabilities of G-AI.
Twenty out of all participants confirmed the efficiency brought by the utilization of G-AI in their domain.
P11, a poster designer, who uses stable diffusion in their workflow to generate new designs for their poster, described the change in efficiency in their experience:
\begin{quote}
    \textit{"I can complete ten times more tasks in a day compared to before. All I have to do is to type down the description of the layout I want and some details, of course. I get the generated drafts and keep refining them by telling the [generative] AI what I need in detail. I have only minor work to do after that."} - P11
\end{quote}

P11 described a common scenario where artists are liberated from the tedium of initializing and iterating over new artistic creations.
The generative power of G-AI fulfills the requirement for completing low-level tasks in a more efficient way than human efforts.

On the reasons for such efficiency, P11 spoke:
\begin{quote}
    \textit{"My work is mainly about switching among designs until a proper layout fits the customers' requirements. I used to have to spend too much time making layouts that neither I nor my customers are sure [if we like or not], now I can use [generative] AI to do this for me in a short time while I can focus on the actual design part of my product."} - P11
\end{quote}
According to P11, G-AI accomplishes variations in design for him with much more efficiency.
We see similar affirmations of variations among most participants.
For instance, P6, a music producer who utilizes GAN-based software to generate soundtracks for his colleague songwriters, speaks on how G-AI assists them in obtaining the variations:
\begin{quote}
    \textit{"Back in the day, we needed to compose the soundtracks one by one or piece by piece to check with my fellas to see if they fit [with the songs]. Now I just need to do one and give it to the software and it will generate so many similar ones. I can choose to modify them all at once with just some text... It changed my work entirely."} - P6
\end{quote}

We see the potential of G-AI in generating variations of artworks, which help artists with the process of creation that requires repeated trials, selections, or iterations of work.
Besides handling variations, collective ideation is also reported to be one of G-AI's tricks for efficiency according to our participants.
E.g., P17, a science fiction writer, highlighted their ideation process with ChatGPT:
\begin{quote}
    \textit{"...It knows how other people write their story. Sometimes I ask it how I should handle the plots in my fiction, and it lists similar plots in others' stories... It sometimes writes its own [story]. I just read them all and see whatever pops out in my mind."} - P17
\end{quote}

The experience of P17 showcased a successful collective ideation with G-AI, where G-AI generates a summarization of other artists' ideas of a similar artistic creation, and the artist/user reviews the set of ideas and refines their own.

Participants confirmed that working with G-AI greatly increased their efficiency in artistic work, specifically in terms of the process of variation trial and idea generation, both of which are critical stages in artistic creation.
This efficiency emerges in various domains of art as long as they require collective ideation or repeated refinery of components.


\subsubsection{Competition with and without G-AI}
With the efficiency brought into artistic creation, there appears a significant gap among artists competing with and without G-AI.
Seven participants reported increases in their profits gained from artistic work with G-AI.
P3 a freelance painter working in the Manga community, described how the chances for contracts increased thanks to G-AI:
\begin{quote}
    \textit{"I get more invitations to draw [Manga] than before now. With Generative AI I can draw [Manga] for different studios. I can draw in different styles... I am not saying I draw better now than before, but I make more for sure."} - P3
\end{quote}

P3's experience depicts scenarios where freelance artists obtain more chances than before to get contracts for certain artwork.
G-AI opens the highway for artists as its users to higher productivity and wider ranges of creation.
It also sets higher standards and expectations from the consumers of art across the domain:
\begin{quote}
    \textit{"Buyers used to invite me to create a certain style of Manga or join a certain studio because I am good at it. Now they just hire me because I can adapt multiple styles quickly."} - P3
\end{quote}

For P3, the deployment of G-AI has shifted the expectation of his work from the consumer community higher, in terms of style adaption.
We see similar circumstances from participants in other art domains, where expertise in multiple styles and variations is preferred.

However, the size of the market never expands due to G-AI.
With consumers satisfied by the co-creation of G-AI and some artists, the remainder of the artists who do not always or at all utilize G-AI are confronted with a crisis of chance.
E.g., P21 as a digital designer who majors in designing widget layouts for e-business elaborated on the cause of his recent unemployment from his former employer:
\begin{quote}
    \textit{"They didn't need me, unfortunately... One of my colleagues showed them how he could [with the help of G-AI] quickly generate hundreds of layout designs with just the names of the merchandise. He iterated over the designs by just typing his needs... I was just working on a short contract. They soon decided they did not need me and let me go afterward... I wish I had known how to do the same thing."} - P21
\end{quote}

The unfortunate experience from P21 showcased the other side of the coin. 
While no extra opportunities for work are created, the utilization of G-AI grants its users advantages over the non-user artists in most domains we identified in our interviews.
(One exception is street art (P19, P20), where no solid market exists and physical work is needed such as graffiti and murals thus exceeding the capabilities of G-AI.)
These advantages of competitors with G-AI over those without it have changed the ecology of the art industry by initiating an unfair competition that does not necessarily depend on competitors' artistic skills and talents - at least not on those by conventional definitions.

\subsubsection{Compromises and the dilemma}
What consequences do these advantages in competition in the art industry lead to?
With this new question arising, we dug into the superficial discussions on the changes in artists' careers.

As was aforementioned, success in the competition induced by G-AI is not necessarily determined by one's artistic skills and talents.
Nevertheless, according to our deeper investigation, it is rather dominated by the usage of G-AI, i.e., artistic works with G-AI overwhelm those without it, regardless of the artistic attributes or aesthetics they may contain.
Conveyed in the competition, this counter-intuitive rationale has compelled artists to yield to the usage of G-AI.
On their motivations for using G-AI, eighteen of the participants said that they were using G-AI mostly in order not to fail in competing against their peers, four posited G-AI as merely a fun tool to obtain ideation, inspiration, or to create with, and only two collaborated with G-AI in pursuit of higher quality of artwork in their disciplines.

P10, a photographer, noted their experience in collaborating with G-AI to create landscape photographs for a magazine:
\begin{quote}
    \textit{"I do not like AI-generated photos, you know. They can sometimes be fake. Imagine a sunset above a lake. You will never expect a reflection of the moon in the lake, but you may see it in a generated photo... I have always thought photography is the art of capturing reality, but they [magazines or audience] don't seem to care about the reality part anymore. They asked me to regenerate more samples with my photo using some AI software. I thought it was gonna be like Photoshop but it wasn't. The generated photos didn't look real to me but they took one anyway... It was a quick call, if I had not got it done in time I might have lost it."} - P10
\end{quote}

For P10, they faced the conflict between his pursuit of realistic photographs and consumers' tastes for different styles.
P10 did not and will not have the same bandwidth as G-AI to experiment with different styles of the same photograph in a short time.
Nevertheless, the chance is either his at that very moment or an easy grasp of other competing photographers who would give in to G-AI.
Under this circumstance, P10 compromised his experience as a user of the G-AI in order to survive the competition with other photographers, regardless of the quality of the photograph generated by the G-AI.

Showcases of similar compromises emerged among other participants.
Painters (P13, P15, and P16) and Mangakas (P1-P5) in our interviews expressed their concerns about possible plagiarism from appropriations of their products as the training data of G-AI models.
P5 brought up their experience of their work being plagiarized by style-transferring G-AI.

An AI-generated drawing is accused by him of \textit{"copying directly my painting in a different style"}.
P7, a music producer, reported that he used G-AI to generate variations of soundtracks with a poorly designed interface with non-intuitive interactions and incomprehensible terminologies.
Yet, they had no other choice because it was the only software available for the variations they needed.
Moreover, P10 and P2, although from different domains, highlighted the similar differences they experienced between using G-AI with (1) rigorous terms (henceforth, rigorous G-AI) where artists' data will not be shared to train the AI and (2) lenient terms (henceforth, lenient G-AI) where artists' data will be shared.
According to P2, rigorous G-AI \textit{“does not have variations in styles in the Mangaka [as much as]"} lenient G-AI, which they used eventually.
P10 also noted their tendency to lenient G-AI because they can \textit{"...iterate my work referencing others photographs... [and] transfer the light and shadow to my shots..."}.

We conclude that despite the efficiency brought into the artistic workflows, G-AI has left at least one negative impact on the artists' community -
It forced the artists into an N-person prisoner's dilemma, where the artists are left with two choices: either equip themselves with G-AI in the workflow or decline it.
Due to its incomparable capabilities in some procedures of artistic workflows, G-AI puts the decliners in an unfavorable position in their peer competition, where their chances and profits are deprived.
Such consequences are insufferable for any individual who wishes to sustain themselves in the art industry.
Subsequently, the outcome of this dilemma is suboptimal - all artists resort to the usage of G-AI, compromising their user experience or data copyright to survive in the peer competition.

This outcome eliminates the comparisons between current G-AI and human factors through an intrusive deployment of G-AI in the artistic workflows.
It also conceals part of the discussions on the comparisons between different designs and applications of G-AI due to the overwhelming capabilities of G-AI in art.
To this end, we further conclude that this suboptimal outcome is negative towards motivating research in Human-GAI symbiosis, posing obstacles and difficulties on the path of developing better interaction techniques and investigating the potential ethical problems within the scope.


\subsection{A Rashomon Situation}

\subsubsection{Can you understand what I am saying?}
To investigate artists' perspectives on G-AI, we asked about how much knowledge they had of the technology of G-AI.
Given much effort made on the Internet elucidating the theory, development, or implementation of G-AI, twenty-two out of all participants reported that they had at least once been in touch with the educational content of any form about G-AI.
Nevertheless, we see negative responses towards the effectiveness of the educational content among the artists.
Twelve participants admitted that they were unable to understand how G-AI actually worked despite the articles they had read or the videos watched.
P13 commented on their experience in learning about G-AI technology:
\begin{quote}
    \textit{"I tried googling the AI that generates painting. I believe it is called Generative Advance[Adversarial] Network or something. There were a lot [of] blogs about it]... I am not a math person. Much as I want to know how it works, the math terms and notations always give me a headache."} - P13
\end{quote}

The instance of P13 resembles those of other participants.
Their notion of G-AI will not be as comprehensive as that of AI engineers or scientists.
Considering the artists as the users of the G-AI systems, elucidations of mathematics and algorithmic of G-AI are apparently user-unfriendly.
In this context, users do not understand, do not need to understand, and do not want to understand the intricate technical details of G-AI.

P4, a Mangaka from a large comic community, expressed their attitude toward understanding G-AI in their disciplines:
\begin{quote}
    \textit{"I use a GAN-based software to change the style of my work sometimes. I also tried some background-changing AI but it was that good... All I care to know is which buttons to click, what they result in, and maybe some efficient tricks for me to master it... Oh, I also wonder if they use my manga or not. I have run into someone else's work for sure when browsing the styles from the software. Will my work be shown there? I don't know... I guess if they ever used mine they should pay me... I honestly do not want to see the math and equations and stuff. I mean, if it steals others' work, it steals. If it works the way I want, it works. That's all I need to know."} - P4
\end{quote}

Comments from P4 representatively conveyed a voice from the artists that as the users of G-AI, they would simply use it as a black box with clear instructions on the usage rather than the intricate theories or principles.
Can they understand what we are saying in terms of numbers, symbols, and equations?
Most likely, no.
Then, what are to be understood and how?
Driven by this new question, we are further convinced of the necessity of understanding artists' subjective vision of G-AI, to advance in the big picture of Human-GAI symbiosis.

\subsubsection{What is G-AI: from an artists' point of view}
To this end, we seek to enquire not only about artists' understanding of the technology of G-AI but more importantly their subjective perceptions as the users of G-AI.
P12 is a poster designer, whose workflow focuses on designing the layouts of characters, backgrounds, visual effects, and texts on the posters for a stand-up comedian (who is also the employer of P12).
P12 discussed the role G-AI has played in their workflow:
\begin{quote}
    \textit{"I need to design and go through thousands of layouts before settling down. Now I can generate the layouts with AI - that saves a lot of time. Yet I still have to go through them myself. Sometimes some modifications are needed... Well if I have to say, it is definitely not creating art design. I think everybody can do what it does. Those are just combinations of widgets. I think we just need a tool like this to save our time."} - P12
\end{quote}

In the case of P12, G-AI is described as an assistive tool that is capable of performing tedious but simple tasks to speed up artistic creation.
Similarly, P18, a news writer who is working for a local newspaper also described their experience with a writing assistant G-AI as a tool:
\begin{quote}
    \textit{"It auto-corrects my writing... To write a good essay on an event, I just type in what happened and it gives me a set of essays from different perspectives... News writing is about being objective but still holding our own opinions. It is good to see from all those angles it provides... I do not think the essay it gives is good enough for publication though. That is why I always write my own, well using the generated as reference."} - P18
\end{quote}

In P18's description, we see how they use AI-generated content as a reference or source of ideation for their news writing.
However, they draw a line between G-AI as a tool and G-AI as a creator similar to P12.
In their opinion, G-AI serves merely as a tool because it is not able to create content that has as high quality as human-created content.
Thus, the majority of the artistic process is done by the human in this context.

Following the same logic as above, in the domains where generated art can be as good or as close, the role of G-AI is perceived otherwise.
P17, the fiction writer aforementioned, described their relationship with G-AI as collaborators:
\begin{quote}
    \textit{"My work strives for ideation. I read lots of books in science to ground my narratives... Now I consult AI about the science background. It not only gives me knowledge of scientific facts but also suggests how I can compose a story by blending science. I chat with it over the ideas as if it is a collaborator... Sometimes I have it proofread the narratives and give me feedback. It is doing what a real human can do - and probably better."} - P17
\end{quote}

In circumstances similar to that of P17, G-AI is capable of composing artistic content that constitutes a part of or potentially the whole of the final product of the artistic workflow.
In the case of P17, G-AI solidifies the scientific background of fiction and blends science into the storyline provided by the human writer, P17.
With concretized contributions as parts of the eventually presented artwork, G-AI is considered a collaborator in such domains.
P6, the music producers also aforementioned, commented on the nature of G-AI as a collaborator:
\begin{quote}
    \textit{"The soundtracks it generates are as good as what I am able to make, honestly sometimes even better. It amazes me with one piece or two [pieces of soundtracks] that fit perfectly in mine. It does some other styles as well. It is just versatile."} - P6
\end{quote}

We notice how participants affirmed the capability of G-AI to produce \textit{"as good as"} human artists.
Such comparability opens up discussion over the artistic values and aesthetics of G-AI products and earns G-AI recognition as a collaborator among the disciplines.
Yet, it also manifests the capability of G-AI to be a competitor in some confrontational situations.
E.g., P24, a screenwriter, who recently went on a strike~\cite{strike} against the use of G-AI in screenwriting in Hollywood, expressed his concerns regarding the competition between G-AI and human screenwriters over the market:
\begin{quote}
    \textit{"The use of AI has already threatened many of my capable domains, like games, novels, movies... My concerns lie most in the credit assignment. In my opinion, if I collaborate with AI on a task, I should take full credit - You cannot give credit for work to something that is not even a human... If they allow free use of AI in these industries, it will kill us [metaphorically]. Then who is gonna write for them? As far as I know, they still need us to teach the AI, right?"} - P24
\end{quote}

The voice of P24 represents an angle viewing the bramble of conflicting opinions and discussions over G-AI ethics.
P24 suggested that no credit be assigned to G-AI but all to the artists who compose with G-AI.
This is a rational bargain from the artists' perspective, yet an extreme one from a holistic view.
One can easily predict that with no credit assigned to G-AI, artists can compose entirely with G-AI and little originality, which results in serious situations such as plagiarism or deteriorated artwork.

Views on G-AI vary not only between artists and AI developers but also among the artist community.
We conclude that the intricacy of G-AI ethics results from (1) the disparity between AI developers' and artists' perspectives on G-AI and (2) the discrepancy of G-AI capabilities among the artistic disciplines.
Specifically, AI developers focus on algorithm performance and define G-AI ethics by the mathematical facts while artists evaluate their experience with G-AI merely from a user perspective.
Moreover, with different capabilities, G-AI has been assigned different roles in artistic creation.
The definition of G-AI creativity and originality depends on its capability in the context.

To this end, we anticipate further discussion on addressing the disparity between AI developers and artists in pursuit of contextual interaction designs, regulations, and best practices.
Also, we identify a call for a taxonomy of G-AI based on artistic capability, which will motivate the research to further scrutinize G-AI ethical concerns among different disciplines.

\subsection{A Tendency to Oppose}

\subsubsection{Opinions on the misusage of G-AI}
In pursuit of artists' perspectives and expectations of G-AI, we received from participants their reflections on the ethical problems induced by the use of G-AI in their domains.

Participants expressed their worry about G-AI shrinking their market.
Besides the strong opposition from P24 and P21 aforementioned, participants like P1 highlighted their concerns as well:
\begin{quote}
    \textit{"Our craft is rooted in passion, creativity, and years of practice. We pour our hearts into each character, each storyline, and each stroke of the pen. But now, with AI capable of mimicking our style and even inventing new plots, I worry that people might start to prefer the convenience and speed of AI-generated manga over our handcrafted works."} - P1
\end{quote}

As described in the previous subsections, the capabilities of G-AI in the replication and summarization of styles of the artists is one of the reasons for it being increasingly dominating in some domains.
What happens if G-AI dominates all human factors in the domains?
Participants seemed to hold overall negative predictions on the future of art with G-AI, due to their lack of confidence in G-AI's understanding of art and aesthetics and its capability to create aesthetics by the human convention:
\begin{quote}
    \textit{"I don't think it understands art. It is just creating meaningless presentations upon request. It cannot create art as a human can. It cannot seize, observe, expand upon a glimpse of the world - this is, in my opinion, what human artists do."} - P14

    \textit{"If there is no human painter, what is left for AI to learn? Are we all just going to learn to appreciate art that is not made by humans?"} -P16
\end{quote}

Moreover, we identified concerns of participants over potential (and even ongoing) G-AI plagiarism in art as well.
G-AI plagiarism in the form of G-AI potentially stealing, e.g. P5 accused G-AI of copying their work directly:
\begin{quote}
    \textit{"I believe they used my painting vastly in training the AI. You can check how similar in layouts and style the product resembles my work... This is intolerable. If another mangaka does this to me, he[/she] will definitely be sued and potentially banned from some communities. Similarly, G-AI should be banned if plagiarism ever happens."} - P5
\end{quote}

G-AI plagiarism also occurs indirectly:
\begin{quote}
    \textit{"The reference plots it generates for me are sometimes from existing literature. It opens up a way for me to quickly review others' work, but it also leaves me in a dangerous spot where I might copy others' writing without even knowing it... If there is no legal or safe usage, we might just need to stop. Nobody wants to plagiarize, after all, even the plagiarizers."} - P17
\end{quote}

P17's case describes a generalizable situation where indirect plagiarism is caused by implicit learning by G-AI of existing human work.
When the user conducts artistic creation, the credit assignment is missing among the user, G-AI, and the authors of the content used in training G-AI.

In addition, participants also displayed awareness of the other potential illegal usage of G-AI, e.g., fabrications (P18) and identity theft (P8):
\begin{quote}
    \textit{"One of the potentialities I see from this technology is that it can generate news even without any context. In other words, it can generate fake news that may sound authentic. This raises significant concerns of mine about the spread of misinformation and the blurring of lines between factual reporting and fabricated narratives."} - P18

    \textit{"I once saw realistic generated photos of [A famous political figure] getting shot. I thought it was real in the first place. The identity theft brought by G-AI can be detrimental. Nobody wants to see their faces in a wanted poster, pornography, or any piece of fake news."} - P8
\end{quote}

We conclude from the interview that the potential or existing misuse of G-AI poses threats to not only the artists across the domain but also individuals outside the disciplines.
To address this profound social impact of G-AI, further research is called for to investigate the usage of G-AI in art, with artists being both the users and the non-users.

\subsubsection{Who is to blame? G-AI or its users}
While participants saw the harms of the misusage of G-AI across the domains, the majority (seventeen out of twenty-four) of them felt that the technology of G-AI is at fault and should be held responsible for the harm.
The aforementioned P8 shared:
\begin{quote}
    \textit{"I noticed instances where this technology blurs the lines between genuine photography and manufactured content... Those fake but realistic photos are a huge threat not just to photography but to the entire society... It is also unfair that someone can sit in front of a computer and generate photos out of imagination, while I have to go Johnny-on-the-spot for the best shot... The technology needs to be refined to ensure it complements our artistic endeavors without overshadowing them."} - P8
\end{quote}

P8 expressed his concerns about fabrication and unfair competition brought about by the usage of G-AI in photography.
Yet, they blame the technology for placing their industry in jeopardy, while overlooking the users of G-AI who in fact fabricated the photos or competed with unfair advantages.
Similarly, screenwriter on strike, P24, emphasized their opposition to G-AI in their domain:
\begin{quote}
    \textit{"The companies are trying to use [generative] AI to replace us in order to reduce the cost. I have to take a stand for not only myself but also my fellow screenwriters. We need to stand our ground to earn a living... It was trained using our work, it was not as good as what we can do, and yet it is taking our jobs away for simply being cheaper. This is totally unacceptable."} - P24
\end{quote}

We identified similar oppositions to the deployment of G-AI in the domains among our participants.
Cases like P8 display how G-AI is maybe utilized to plagiarize, fabricate, or deluge with different artwork, while cases like P24 manifest how G-AI is seen as a rival in the competition of the job markets.
Due to these ethical problems, artists display a strong tendency to oppose G-AI. 
We argue that from the artists' perspective, this tendency to oppose is rational.
They do not possess the knowledge to comprehend the design goals of interactions with G-AI, supervise the training and deployment of G-AI, or objectively converse about the capabilities and constraints of G-AI.
Moreover, they are not obligated to do so.
Therefore, their subjective perception results in an opposing viewpoint of the technology rather than its usage, i.e. its interaction design.

However, the opposition to G-AI forms a confrontation between the advancement of the technology and its engaging users and non-users.
To not put the promising power of G-AI and the efforts of AI developers to no avail, we call for discussions on how to redirect artists' opposition into propulsion to determine the design rationale for G-AI in art.

\subsection{Summary}
To this end, we have investigated the first question "How are artists’ careers changed by G-AI?" and the second "What are artists’ perspectives on G-AI?".

Through our interviews, we saw the positive impact of G-AI on the artistic workflow, in terms of efficiency and ideation.
Yet, we highlighted the disparity in competition caused by the overwhelming productivity of G-AI, which resulted in a dilemma where artists eventually resort to G-AI, regardless of their user experience.

In pursuit of the answers to artists' perspectives, we identified the gap between artists' and AI developers' understanding of G-AI, implying the vain efforts of educating artists on the theorem or mathematics behind G-AI.
Moreover, we manifested the discrepancy in the opinions of the artists from different disciplines, which is subject to the different capabilities of G-AI in these disciplines.

Then, we identified a tendency to oppose G-AI among artists.
We analyzed the rationale behind this opposition to G-AI technology itself rather than its misusage and proposed to redirect this opposition to push the research on interaction design for G-AI in art.

Based on the results gained from the interviews, we seek the answers to the third question "How do we bridge artists’ perspectives with future research on G-AI?".
We will do this via a thorough discussion in the following section, addressing the problems identified.

\section{Discussion}

In this section, we will spark a discussion on the results we draw from the interviews, identify the major conflicts across communities, uncover the misunderstandings on the techniques, and elucidate the potential effort that can facilitate an alignment between both the artists and generative G-AI designers. 

\subsection{Addressing the Dilemma Between Artists}

The capability of G-AI to generate variations of artwork, especially those that require tedious efforts on trial and selection, such as layout design or sound-tracking composition, has greatly increased human efficiency.
The increase in efficiency, on the one hand, improves the productivity of the artists who leverage G-AI in their work, but on the other hand, it brings a sharp and unbalanced competition between artists with and without AI.
This unbalanced competition results in the problem of the N-person Prisoner’s dilemma, in which artists have to compromise with the use of G-AI to survive in the market, at the cost of their personal user experience and taste in art.
To address this dilemma, we will illuminate the following discussion from both the artist's point of view and the G-AI developer's point of view.

\subsubsection{Embrace the new standard of art}

The introduction of G-AI has had a major impact on the digital creation industry, altering the perspective of creators, artists, and the industry on the standard of art. 
With its high productivity, tasks that used to require a lot of effort can now be completed with a few prompts and selections.
These works are likely to be seen as tasks at the low entry level with fewer demands on the artist's creativity, experience, and skills, thus reducing the value of art. 
Even creators using G-AI will not be as competitive as before and will struggle in the low-value market.
This necessitates the artist and the industry to come to \textbf{a consensus on the new standard of art} to take advantage of G-AI and also to help creators succeed during this G-AI revolution.

\subsubsection{G-AI as Collaborators}

Artists are now viewing G-AI as a powerful tool to gain an edge and remain competitive in the market, rather than as a collaborator or colleague.
They tend to overlook the potential of G-AI to aid creativity, avoid stagnation, and facilitate knowledge transfer.
These various potentials of G-AI can offer more opportunities for artists and enhance their professional abilities, refine their sense of aesthetics, and investigate different styles.
Therefore, it is essential that artists regard G-AI as a partner and investigate \textbf{more uses of G-AI than its ability to compete}.
At the same time, more research and studies are needed to redefine and create new connections between human designers and G-AI.

\subsubsection{Lessens the Skill Barrier}

The user described the situation of unemployment from his former employer due to the lack of G-AI skills.
We anticipate that this will become more common for creators who have not been exposed to G-AI due to financial, Internet, or computational resources.
Additionally, there is a disparity in learning ability among creators, with experienced professionals more accustomed to existing tools and processes.
Some business training organizations offer additional G-AI skill training to those with budgets, giving them an advantage over those who are not.
The difference in G-AI skills leads to an unequal competitive landscape in the digital art market, with skilled creators earning more and those without the necessary skills becoming unemployed.
To ensure that the benefits of G-AI are available to the majority of the community, rather than just a select few, steps must be taken to \textbf{reduce the skill barrier of G-AI}.

\subsubsection{Propose New Ideology of Art}

G-AI, on the one hand, presents challenges for existing professionals, and, on the other hand, offers opportunities to novices.
By lowering the entry barrier to this field, people without professional training can express their creativity.
Furthermore, the G-AI's capability of generation provides a wide range of variations, from low-level elements such as shapes and colors to high-level representations such as layouts and styles, which offers creators the chance to explore the vast potential of creativity and \textbf{discover new aesthetics that may not be rooted in human history or culture}.
As human creativity can be unleashed and nurtured by this powerful tool, new ideologies of art are needed to discuss potential ways of harnessing this change.

\subsubsection{User Experience Over Efficiency}

In order to make G-AI more convincing, developers tend to focus on improving its efficiency rather than user experience design.
Most systems lack consideration of human factors, such as interaction design, human-in-the-loop collaboration, and explainability.
These considerations are essential for users to recognize the advantages and limitations of G-AI.
Without them, users may be forced to accept the technology with a passive and doubtful attitude, leading to opposition to the technique and not a lack of human-centered system design.
From the perspective of the G-AI developer, it is necessary to \textbf{prioritize user experience over efficiency}.
By providing a better user experience to creators, we can prevent G-AI from becoming the terminator of human art and creativity.




\subsection{Bridging the differences between Artist and G-AI}
\subsubsection{Understanding G-AI Principles} 
We have identified an issue within the creative art and design community: artists often lack a deep understanding of G-AI models, making their utilization increasingly challenging.
The complexity of G-AI systems has created a barrier due to artists' limited knowledge of their benefits and applications.
To address this challenge, it is crucial to provide accessible and user-friendly resources.
Effective use of G-AI models can be facilitated through the creation of clear, comprehensive documentation and tutorials.
These materials should focus on \textbf{explaining functionality and limitations in a straightforward manner}, without delving into mathematical derivations and formulas.
Some of the ways in which developers and creators can create such documentation are by providing step-by-step tutorials about usage, more visual and interactive platforms, some examples and use cases, and updates about the latest developments.
By adopting this approach, we can significantly enhance artists' ability to work effectively with G-AI systems, encouraging innovation and creativity across various artistic domains. 

\subsubsection{Design space of interaction with G-AI}
To fully harness the potential of G-AI systems and models, seamless interaction between humans and these systems is imperative.
Some artists have highlighted the existing limitations of such systems \textit{"... does not possess a desired degree of freedom in creating ... (P4)"}, emphasizing the importance of considering the artist's viewpoint during the initial design phase and tailoring the interaction with G-AI systems to their creative needs.
Often, developers prioritize utility by concentrating more on developing an AI to make the system work rather than usability.
In order to create a more artist-centric system, it is crucial to incorporate their insights and preferences.
We envision that conducting a comprehensive user study on design principles from the artist's perspective is essential.
Further research in developing \textbf{a taxonomy spanning the dimensions of G-AI based on artistic capabilities} is much needed. This will align the capabilities of the G-AI systems with the practical needs and imaginative aspirations of artists.
Furthermore, These designs should be dynamic, evolving over time based on ongoing artist input and suggestions. This iterative approach ensures that the G-AI system becomes a tool that artists can effectively utilize and engage with in their creative processes. 

\subsubsection{Artist in the loop G-AI}
We found mixed opinions regarding the usage of G-AI systems in the same domain.
While certain artists expressed contentment with the creative outputs produced by these AI systems, others conveyed dissatisfaction with the generated content.
This shows the need for \textbf{customization of design for the use of G-AI varying with the artist}.
This divergence underscores the imperative for a tailored approach in adapting G-AI designs to individual artists' preferences and needs.
Addressing this requires the proactive incorporation of feedback, insights, and artistic inclinations from creators.
By doing so, developers can instill a heightened level of flexibility and personalization within G-AI systems, thereby fostering an environment conducive to unfettered artistic expression.

\subsubsection{Explainability}
A different user brought to light the challenge of inconsistent outcomes arising from the same prompt, \textit{"...there is some randomness in the application. The results varied with the same input ... I wonder how it got those pictures.(P8)"}
This discrepancy underscores the deficiency in explainability inherent to G-AI systems.
To mitigate this issue, AI developers should formulate \textbf{visual representation methods that facilitate artists' comprehension} of the intricate relationship between inputs and outputs.
Additionally, it is imperative to incorporate controlled content generation tools that empower users to manipulate and govern the resultant creations.
This multifaceted approach will not only empower artists with a sense of agency but also facilitate a comprehensive understanding of the mechanisms driving the G-AI system's creative output.

\subsubsection{Artist G-AI collaborations}
We discovered that certain artists expressed dissatisfaction and opted not to utilize the tool due to the AI module's unhelpful output.
One plausible explanation is that the G-AI system fails to grasp the intended nuances of human input, a vital aspect of effective collaborations.
The results produced by G-AI models should be linked to the artist's context.
Achieving \textbf{a blend of the artist's intention and contextual factors} fosters a collaborative environment leading to well-informed choices.
This collaboration between the artist and G-AI system emerges \textbf{not just as a tool, but as a creative partner} that empowers artists to push the boundaries of their imagination while seamlessly integrating cutting-edge technology.

\subsection{Exploring Social Aspects}

\subsubsection{Unemployment Fear}
We discovered that numerous artists expressed concerns about potential unemployment due to the rise of G-AI models taking over their roles.
While G-AI has demonstrated its capability to produce intricately detailed art, surpassing even the skills of elite artists, it also exhibits weaknesses, such as its inability to consistently render realistic human hands (resulting in unusual finger bending).
This underscores the reality that despite the widespread adoption of G-AI, it is not without its drawbacks and still artists have a crucial role to play.
We suggest comprehensive research aimed at \textbf{probing the capacities and limitations of G-AI within the realm of art}.
Such an initiative would provide valuable insights to artists, enabling them to discern the domains in which G-AI can offer assistance for an increase in productivity and those where its utilization would be less viable.
Additionally, it will also help artists \textbf{identify the novel employment prospects caused by G-AI}.

\subsubsection{Plagiarism and Privacy}
Plagiarism is a concern when the output of AI-generated content closely resembles existing works, including those created by artists. 
These G-AI models are trained using datasets gathered from the internet, inadvertently leading to outputs that produce output similar to the existing arts. 
Artists have shown concerns about similar designs generated by G-AI.
To tackle this issue, a suggested approach involves employing a "machine unlearning" technique.
This technique would allow developers to selectively \textbf{remove specific samples from the training data} without necessitating a complete retraining of the model. 
Furthermore, transparency should be maintained by AI developers to provide users with insights into the technology's inner workings, training processes, and potential limitations.
By fostering a clear understanding of the capabilities and constraints of AI systems, users can engage with the technology more responsibly and with awareness.

\subsubsection{Ethical concerns}
Every technology comes with both benefits and drawbacks.
G-AI has indeed improved various fields, including art, by enhancing efficiency and quality.
However, it has also opened the door to potentially fraudulent activities.
One of the users pointed out the wrong use of generated content in winning the competition (P21). 
Furthermore, Deepfakes and manipulated media can indeed have negative implications for the integrity and authenticity of art.
The use of generative output detection and proper credit assignment can certainly be valuable measures to address these issues.
\textbf{Generative Output Detection:} Implementing techniques to identify whether a piece of art or content has been generated by AI can help distinguish between original and AI-assisted work.
\textbf{Credit Assignment:} Properly attributing the contribution of both the artist and the G-AI model is crucial. Clear disclosure of AI involvement ensures that the audience is aware of the level of human input and AI assistance. 
By combining these approaches, there may be potentially reduced fraudulent activities in art competitions and other contexts where AI-generated content is involved.
It's important to strike a balance between utilizing AI for creative enhancement and maintaining the integrity of artistic creation. Further research should be done to provide more solutions to stop fraudulent activities.  

\section{Limitations}



We interviewed a wide variety of participants from different disciplines of art and investigated their subjective perspectives on G-AI in art.
However, there exist several aspects of our study design that limit our findings.
We outline these limitations and explore how they might be addressed in future work.

One of the limitations of our work is that we focused on a limited set of art.
Art is known for its interdisciplinary nature, often transcending boundaries and merging various forms.
In order to address the impact of G-AI on art, we concentrated only on the categories of art in which G-AI meets a minimum bar of the capability to contribute to artistic creation.
Such categories include visual art, music, literature, film, architecture, and digital art.
We excluded categories of art that were not affected subject the current capabilities of G-AI, such as martial arts, physical sculpture, abstract expressionism, handicrafts, and performance art.
These categories by nature require physical movement or expression in reality and are thus not yet as affected as the aforementioned.
However, we do not underestimate the aligning artistic nature they share with the included categories.
By concentrating exclusively on specific categories, there's a risk of neglecting the cross-pollination of ideas, techniques, and influences that occur between diverse art forms.
This omission could hinder the exploration of how art evolves and interacts within a broader cultural and artistic landscape.
Further research is anticipated to deeply investigate how G-AI can affect the art that requires physical components, with these constraints.
Another aspect of this limitation is that the categories we have located might not be fully covered by our participants.
Even with the best intentions and investments in our recruitment, we cannot claim that our volume of participants is able to cover the broad topic of art.

A second limitation of our work lies in our recruitment methodology.
When participants are recruited most via the Internet, the sample may skew towards individuals who are more technologically inclined, digitally savvy, or have a pre-existing interest in generative art.
This bias can lead to a lack of representation from individuals who may not have regular access to online platforms or those who hold different viewpoints.
Consequently, the insights gathered from this group may not accurately reflect the broader population's attitudes and opinions on generative art.
Furthermore, the online environment can also foster a self-selection bias.
Participants who actively choose to engage in online discussions about generative art may possess a greater degree of familiarity with the subject matter, potentially resulting in more informed opinions.
This bias could inadvertently exclude individuals who are less knowledgeable about generative art but could provide valuable perspectives.
Yet, we argue that the goal of our interviews is to obtain subjective feedback from the artist community to foster a propulsion to the HCI research on G-AI in art.
We hold dear the fiercely outstanding and subject voices of the artists.
We anticipate future research on both subjective feedback and objective evaluations in the context of G-AI in art, yet we persist that this limitation will not undermine the calls for HCI research that are urgently to be addressed in this paper.
\section{Conclusion}

In this paper, we contribute, to the best of our knowledge, the first interview study that focuses on the human-computer interaction between artists and the emerging technology of G-AI.
Aiming to understand the artists' understanding and expectations of G-AI as well as the impact of G-AI on their careers, we conducted semi-structured interviews with twenty-five artists.
Through our interviews, we discovered a dilemma the artists are put in by the technology of G-AI, a disparity in the understanding of G-AI between the artists and the developers, and a tendency to oppose the use of G-AI.
To this end, we identify gaps in the research of HCI between artists and G-AI.
From the G-AI point of view, the development of interaction design needs to pace up with the exploding technology to grant user-friendly interactions.
From the artists' point of view, the capabilities of G-AI should be appreciated, and a new ideology of aesthetics and a new standard of artistic ability are to be embraced.
From the societal point of view, ethical problems such as credit assignment are yet to be addressed.
We discuss the open questions on how to tackle the challenges in the views above.
We envision this work to provide a valuable reference to future research on the missing piece of HCI in the topic of G-AI in art.

\begin{acks}
To all readers, for reading my article.
\end{acks}

\bibliographystyle{ACM-Reference-Format}
\bibliography{main}


\begin{thebibliography}{116}


\ifx \showCODEN    \undefined \def \showCODEN     #1{\unskip}     \fi
\ifx \showDOI      \undefined \def \showDOI       #1{#1}\fi
\ifx \showISBNx    \undefined \def \showISBNx     #1{\unskip}     \fi
\ifx \showISBNxiii \undefined \def \showISBNxiii  #1{\unskip}     \fi
\ifx \showISSN     \undefined \def \showISSN      #1{\unskip}     \fi
\ifx \showLCCN     \undefined \def \showLCCN      #1{\unskip}     \fi
\ifx \shownote     \undefined \def \shownote      #1{#1}          \fi
\ifx \showarticletitle \undefined \def \showarticletitle #1{#1}   \fi
\ifx \showURL      \undefined \def \showURL       {\relax}        \fi
\providecommand\bibfield[2]{#2}
\providecommand\bibinfo[2]{#2}
\providecommand\natexlab[1]{#1}
\providecommand\showeprint[2][]{arXiv:#2}

\bibitem[deb(2023a)]%
        {debleepost}
 \bibinfo{year}{2023}\natexlab{a}.
\newblock \bibinfo{title}{{Deb Lee's post on Social Media on the Plagiarism of
  G-AI on their artwork}}.
\newblock
\newblock
\urldef\tempurl%
\url{https://x.com/jdebbiel/status/1601678889301905408?s=20}
\showURL{%
\tempurl}
\newblock
\shownote{Accessed: 2023/09/14}.


\bibitem[lif(2023a)]%
        {liflagpost}
 \bibinfo{year}{2023}\natexlab{a}.
\newblock \bibinfo{title}{{Li Flag's original post on Social Media of their
  artwork}}.
\newblock
\newblock
\urldef\tempurl%
\url{https://x.com/jdebbiel/status/1601678889301905408?s=20}
\showURL{%
\tempurl}
\newblock
\shownote{Accessed: 2023/09/14}.


\bibitem[mid(2023)]%
        {midjourneywebsite}
 \bibinfo{year}{2023}\natexlab{}.
\newblock \bibinfo{title}{{Midjourney}}.
\newblock
\newblock
\urldef\tempurl%
\url{https://www.midjourney.com/}
\showURL{%
\tempurl}
\newblock
\shownote{Accessed: 2023/08/02}.


\bibitem[ope(2023)]%
        {openaiwebsite}
 \bibinfo{year}{2023}\natexlab{}.
\newblock \bibinfo{title}{OpenAI}.
\newblock
\newblock
\urldef\tempurl%
\url{https://www.openai.com/}
\showURL{%
\tempurl}
\newblock
\shownote{Accessed: 2023/08/02}.


\bibitem[man(2023)]%
        {manifesto}
 \bibinfo{year}{2023}\natexlab{}.
\newblock \bibinfo{title}{Our Manifesto for AI companies regulation in Europe}.
\newblock
\newblock
\urldef\tempurl%
\url{https://www.egair.eu/#manifesto}
\showURL{%
\tempurl}
\newblock
\shownote{Last accessed on August 2, 2023}.


\bibitem[deb(2023b)]%
        {debleewebsite}
 \bibinfo{year}{2023}\natexlab{b}.
\newblock \bibinfo{title}{{The personal website of Deb Lee}}.
\newblock
\newblock
\urldef\tempurl%
\url{https://debleeart.com/}
\showURL{%
\tempurl}
\newblock
\shownote{Accessed: 2023/09/14}.


\bibitem[lif(2023b)]%
        {liflagwebsite}
 \bibinfo{year}{2023}\natexlab{b}.
\newblock \bibinfo{title}{{The personal website of Li Flag}}.
\newblock
\newblock
\urldef\tempurl%
\url{https://www.pixiv.net/users/14165905}
\showURL{%
\tempurl}
\newblock
\shownote{Accessed: 2023/09/14}.


\bibitem[Abd-Elaal et~al\mbox{.}(2019)]%
        {abd2019artificial}
\bibfield{author}{\bibinfo{person}{El-Sayed Abd-Elaal}, \bibinfo{person}{SH
  Gamage}, \bibinfo{person}{Julie~E Mills}, {et~al\mbox{.}}}
  \bibinfo{year}{2019}\natexlab{}.
\newblock \showarticletitle{Artificial intelligence is a tool for cheating
  academic integrity}. In \bibinfo{booktitle}{\emph{30th Annual Conference for
  the Australasian Association for Engineering Education (AAEE 2019): Educators
  becoming agents of change: Innovate, integrate, motivate}}.
  \bibinfo{pages}{397--403}.
\newblock


\bibitem[Abrahamsen and Yao(2023)]%
        {abrahamsen2023inventing}
\bibfield{author}{\bibinfo{person}{Nilin Abrahamsen} {and}
  \bibinfo{person}{Jiahao Yao}.} \bibinfo{year}{2023}\natexlab{}.
\newblock \bibinfo{title}{Inventing painting styles through natural
  inspiration}.
\newblock
\newblock
\showeprint[arxiv]{2305.12015}~[cs.CV]


\bibitem[Adams(2022)]%
        {adams2022one}
\bibfield{author}{\bibinfo{person}{Jane Adams}.}
  \bibinfo{year}{2022}\natexlab{}.
\newblock \showarticletitle{One Artist's Personal Reflections on Methods and
  Ethics of Creating Mixed Media Artificial Intelligence Art}.
\newblock \bibinfo{journal}{\emph{arXiv preprint arXiv:2212.11232}}
  (\bibinfo{year}{2022}).
\newblock


\bibitem[AIArtists(2023)]%
        {AIArtists}
\bibfield{author}{\bibinfo{person}{AIArtists}.}
  \bibinfo{year}{2023}\natexlab{}.
\newblock \bibinfo{title}{AIArtists.org}.
\newblock
\newblock
\urldef\tempurl%
\url{https://aiartists.org/}
\showURL{%
\tempurl}
\newblock
\shownote{Last accessed on August 2, 2023}.


\bibitem[Ajani(2022)]%
        {ajani2022human}
\bibfield{author}{\bibinfo{person}{Gianmaria Ajani}.}
  \bibinfo{year}{2022}\natexlab{}.
\newblock \showarticletitle{Human Authorship and Art Created by Artificial
  Intelligence--Where Do We Stand?}. In \bibinfo{booktitle}{\emph{Digital
  Ethics}}. Nomos Verlagsgesellschaft mbH \& Co. KG, \bibinfo{pages}{253--270}.
\newblock


\bibitem[Al-Rubaie and Chang(2019)]%
        {al2019privacy}
\bibfield{author}{\bibinfo{person}{Mohammad Al-Rubaie} {and}
  \bibinfo{person}{J~Morris Chang}.} \bibinfo{year}{2019}\natexlab{}.
\newblock \showarticletitle{Privacy-preserving machine learning: Threats and
  solutions}.
\newblock \bibinfo{journal}{\emph{IEEE Security \& Privacy}}
  \bibinfo{volume}{17}, \bibinfo{number}{2} (\bibinfo{year}{2019}),
  \bibinfo{pages}{49--58}.
\newblock


\bibitem[Ali et~al\mbox{.}(2023)]%
        {ali2023constructing}
\bibfield{author}{\bibinfo{person}{Safinah Ali}, \bibinfo{person}{Daniella
  DiPaola}, \bibinfo{person}{Randi Williams}, \bibinfo{person}{Prerna Ravi},
  {and} \bibinfo{person}{Cynthia Breazeal}.} \bibinfo{year}{2023}\natexlab{}.
\newblock \showarticletitle{Constructing Dreams using Generative AI}.
\newblock \bibinfo{journal}{\emph{arXiv preprint arXiv:2305.12013}}
  (\bibinfo{year}{2023}).
\newblock


\bibitem[Baio(2023)]%
        {Baio2023Invasive}
\bibfield{author}{\bibinfo{person}{Andy Baio}.}
  \bibinfo{year}{2023}\natexlab{}.
\newblock \bibinfo{booktitle}{\emph{Invasive Diffusion: How one unwilling
  illustrator found herself turned into an AI model}}.
\newblock
\urldef\tempurl%
\url{https://waxy.org/2022/11/invasive-diffusion-how-one-unwilling-illustrator-found-herself-turned-into-an-ai-model/}
\showURL{%
\tempurl}


\bibitem[Bellaiche et~al\mbox{.}(2023)]%
        {bellaiche2023humans}
\bibfield{author}{\bibinfo{person}{Lucas Bellaiche}, \bibinfo{person}{Rohin
  Shahi}, \bibinfo{person}{Martin~Harry Turpin}, \bibinfo{person}{Anya
  Ragnhildstveit}, \bibinfo{person}{Shawn Sprockett},
  \bibinfo{person}{Nathaniel Barr}, \bibinfo{person}{Alexander Christensen},
  {and} \bibinfo{person}{Paul Seli}.} \bibinfo{year}{2023}\natexlab{}.
\newblock \showarticletitle{Humans versus AI: whether and why we prefer
  human-created compared to AI-created artwork}.
\newblock \bibinfo{journal}{\emph{Cognitive Research: Principles and
  Implications}} \bibinfo{volume}{8}, \bibinfo{number}{1}
  (\bibinfo{year}{2023}), \bibinfo{pages}{1--22}.
\newblock


\bibitem[Bird et~al\mbox{.}(2023)]%
        {bird2023typology}
\bibfield{author}{\bibinfo{person}{Charlotte Bird}, \bibinfo{person}{Eddie~L
  Ungless}, {and} \bibinfo{person}{Atoosa Kasirzadeh}.}
  \bibinfo{year}{2023}\natexlab{}.
\newblock \showarticletitle{Typology of Risks of Generative Text-to-Image
  Models}.
\newblock \bibinfo{journal}{\emph{arXiv preprint arXiv:2307.05543}}
  (\bibinfo{year}{2023}).
\newblock


\bibitem[Boden and Edmonds(2009)]%
        {boden2009generative}
\bibfield{author}{\bibinfo{person}{Margaret~A Boden} {and}
  \bibinfo{person}{Ernest~A Edmonds}.} \bibinfo{year}{2009}\natexlab{}.
\newblock \showarticletitle{What is generative art?}
\newblock \bibinfo{journal}{\emph{Digital Creativity}} \bibinfo{volume}{20},
  \bibinfo{number}{1-2} (\bibinfo{year}{2009}), \bibinfo{pages}{21--46}.
\newblock


\bibitem[Bommasani et~al\mbox{.}(2022)]%
        {bommasani2022opportunities}
\bibfield{author}{\bibinfo{person}{Rishi Bommasani}, \bibinfo{person}{Drew~A.
  Hudson}, \bibinfo{person}{Ehsan Adeli}, \bibinfo{person}{Russ Altman},
  \bibinfo{person}{Simran Arora}, \bibinfo{person}{Sydney von Arx},
  \bibinfo{person}{Michael~S. Bernstein}, \bibinfo{person}{Jeannette Bohg},
  \bibinfo{person}{Antoine Bosselut}, \bibinfo{person}{Emma Brunskill},
  \bibinfo{person}{Erik Brynjolfsson}, \bibinfo{person}{Shyamal Buch},
  \bibinfo{person}{Dallas Card}, \bibinfo{person}{Rodrigo Castellon},
  \bibinfo{person}{Niladri Chatterji}, \bibinfo{person}{Annie Chen},
  \bibinfo{person}{Kathleen Creel}, \bibinfo{person}{Jared~Quincy Davis},
  \bibinfo{person}{Dora Demszky}, \bibinfo{person}{Chris Donahue},
  \bibinfo{person}{Moussa Doumbouya}, \bibinfo{person}{Esin Durmus},
  \bibinfo{person}{Stefano Ermon}, \bibinfo{person}{John Etchemendy},
  \bibinfo{person}{Kawin Ethayarajh}, \bibinfo{person}{Li Fei-Fei},
  \bibinfo{person}{Chelsea Finn}, \bibinfo{person}{Trevor Gale},
  \bibinfo{person}{Lauren Gillespie}, \bibinfo{person}{Karan Goel},
  \bibinfo{person}{Noah Goodman}, \bibinfo{person}{Shelby Grossman},
  \bibinfo{person}{Neel Guha}, \bibinfo{person}{Tatsunori Hashimoto},
  \bibinfo{person}{Peter Henderson}, \bibinfo{person}{John Hewitt},
  \bibinfo{person}{Daniel~E. Ho}, \bibinfo{person}{Jenny Hong},
  \bibinfo{person}{Kyle Hsu}, \bibinfo{person}{Jing Huang},
  \bibinfo{person}{Thomas Icard}, \bibinfo{person}{Saahil Jain},
  \bibinfo{person}{Dan Jurafsky}, \bibinfo{person}{Pratyusha Kalluri},
  \bibinfo{person}{Siddharth Karamcheti}, \bibinfo{person}{Geoff Keeling},
  \bibinfo{person}{Fereshte Khani}, \bibinfo{person}{Omar Khattab},
  \bibinfo{person}{Pang~Wei Koh}, \bibinfo{person}{Mark Krass},
  \bibinfo{person}{Ranjay Krishna}, \bibinfo{person}{Rohith Kuditipudi},
  \bibinfo{person}{Ananya Kumar}, \bibinfo{person}{Faisal Ladhak},
  \bibinfo{person}{Mina Lee}, \bibinfo{person}{Tony Lee}, \bibinfo{person}{Jure
  Leskovec}, \bibinfo{person}{Isabelle Levent}, \bibinfo{person}{Xiang~Lisa
  Li}, \bibinfo{person}{Xuechen Li}, \bibinfo{person}{Tengyu Ma},
  \bibinfo{person}{Ali Malik}, \bibinfo{person}{Christopher~D. Manning},
  \bibinfo{person}{Suvir Mirchandani}, \bibinfo{person}{Eric Mitchell},
  \bibinfo{person}{Zanele Munyikwa}, \bibinfo{person}{Suraj Nair},
  \bibinfo{person}{Avanika Narayan}, \bibinfo{person}{Deepak Narayanan},
  \bibinfo{person}{Ben Newman}, \bibinfo{person}{Allen Nie},
  \bibinfo{person}{Juan~Carlos Niebles}, \bibinfo{person}{Hamed Nilforoshan},
  \bibinfo{person}{Julian Nyarko}, \bibinfo{person}{Giray Ogut},
  \bibinfo{person}{Laurel Orr}, \bibinfo{person}{Isabel Papadimitriou},
  \bibinfo{person}{Joon~Sung Park}, \bibinfo{person}{Chris Piech},
  \bibinfo{person}{Eva Portelance}, \bibinfo{person}{Christopher Potts},
  \bibinfo{person}{Aditi Raghunathan}, \bibinfo{person}{Rob Reich},
  \bibinfo{person}{Hongyu Ren}, \bibinfo{person}{Frieda Rong},
  \bibinfo{person}{Yusuf Roohani}, \bibinfo{person}{Camilo Ruiz},
  \bibinfo{person}{Jack Ryan}, \bibinfo{person}{Christopher Ré},
  \bibinfo{person}{Dorsa Sadigh}, \bibinfo{person}{Shiori Sagawa},
  \bibinfo{person}{Keshav Santhanam}, \bibinfo{person}{Andy Shih},
  \bibinfo{person}{Krishnan Srinivasan}, \bibinfo{person}{Alex Tamkin},
  \bibinfo{person}{Rohan Taori}, \bibinfo{person}{Armin~W. Thomas},
  \bibinfo{person}{Florian Tramèr}, \bibinfo{person}{Rose~E. Wang},
  \bibinfo{person}{William Wang}, \bibinfo{person}{Bohan Wu},
  \bibinfo{person}{Jiajun Wu}, \bibinfo{person}{Yuhuai Wu},
  \bibinfo{person}{Sang~Michael Xie}, \bibinfo{person}{Michihiro Yasunaga},
  \bibinfo{person}{Jiaxuan You}, \bibinfo{person}{Matei Zaharia},
  \bibinfo{person}{Michael Zhang}, \bibinfo{person}{Tianyi Zhang},
  \bibinfo{person}{Xikun Zhang}, \bibinfo{person}{Yuhui Zhang},
  \bibinfo{person}{Lucia Zheng}, \bibinfo{person}{Kaitlyn Zhou}, {and}
  \bibinfo{person}{Percy Liang}.} \bibinfo{year}{2022}\natexlab{}.
\newblock \bibinfo{title}{On the Opportunities and Risks of Foundation Models}.
\newblock
\newblock
\showeprint[arxiv]{2108.07258}~[cs.LG]


\bibitem[Bran et~al\mbox{.}(2023)]%
        {bran2023emerging}
\bibfield{author}{\bibinfo{person}{Emanuela Bran}, \bibinfo{person}{Cosima
  Rughini}, \bibinfo{person}{Gheorghe Nadoleanu}, {and}
  \bibinfo{person}{Michael~G Flaherty}.} \bibinfo{year}{2023}\natexlab{}.
\newblock \showarticletitle{The Emerging Social Status of Generative AI:
  Vocabularies of AI Competence in Public Discourse}. In
  \bibinfo{booktitle}{\emph{Proceedings of The 24th Conference on Control
  Systems and Computer Science (CSCS24 2023). IEEE}}.
\newblock


\bibitem[Brock et~al\mbox{.}(2018)]%
        {brock2018large}
\bibfield{author}{\bibinfo{person}{Andrew Brock}, \bibinfo{person}{Jeff
  Donahue}, {and} \bibinfo{person}{Karen Simonyan}.}
  \bibinfo{year}{2018}\natexlab{}.
\newblock \showarticletitle{Large scale GAN training for high fidelity natural
  image synthesis}.
\newblock \bibinfo{journal}{\emph{arXiv preprint arXiv:1809.11096}}
  (\bibinfo{year}{2018}).
\newblock


\bibitem[Brown et~al\mbox{.}(2020)]%
        {brown2020language}
\bibfield{author}{\bibinfo{person}{Tom Brown}, \bibinfo{person}{Benjamin Mann},
  \bibinfo{person}{Nick Ryder}, \bibinfo{person}{Melanie Subbiah},
  \bibinfo{person}{Jared~D Kaplan}, \bibinfo{person}{Prafulla Dhariwal},
  \bibinfo{person}{Arvind Neelakantan}, \bibinfo{person}{Pranav Shyam},
  \bibinfo{person}{Girish Sastry}, \bibinfo{person}{Amanda Askell},
  {et~al\mbox{.}}} \bibinfo{year}{2020}\natexlab{}.
\newblock \showarticletitle{Language models are few-shot learners}.
\newblock \bibinfo{journal}{\emph{Advances in neural information processing
  systems}}  \bibinfo{volume}{33} (\bibinfo{year}{2020}),
  \bibinfo{pages}{1877--1901}.
\newblock


\bibitem[Brundage et~al\mbox{.}(2018)]%
        {brundage2018malicious}
\bibfield{author}{\bibinfo{person}{Miles Brundage}, \bibinfo{person}{Shahar
  Avin}, \bibinfo{person}{Jack Clark}, \bibinfo{person}{Helen Toner},
  \bibinfo{person}{Peter Eckersley}, \bibinfo{person}{Ben Garfinkel},
  \bibinfo{person}{Allan Dafoe}, \bibinfo{person}{Paul Scharre},
  \bibinfo{person}{Thomas Zeitzoff}, \bibinfo{person}{Bobby Filar},
  {et~al\mbox{.}}} \bibinfo{year}{2018}\natexlab{}.
\newblock \showarticletitle{The malicious use of artificial intelligence:
  Forecasting, prevention, and mitigation}.
\newblock \bibinfo{journal}{\emph{arXiv preprint arXiv:1802.07228}}
  (\bibinfo{year}{2018}).
\newblock


\bibitem[Brundage et~al\mbox{.}(2020)]%
        {brundage2020toward}
\bibfield{author}{\bibinfo{person}{Miles Brundage}, \bibinfo{person}{Shahar
  Avin}, \bibinfo{person}{Jasmine Wang}, \bibinfo{person}{Haydn Belfield},
  \bibinfo{person}{Gretchen Krueger}, \bibinfo{person}{Gillian Hadfield},
  \bibinfo{person}{Heidy Khlaaf}, \bibinfo{person}{Jingying Yang},
  \bibinfo{person}{Helen Toner}, \bibinfo{person}{Ruth Fong}, {et~al\mbox{.}}}
  \bibinfo{year}{2020}\natexlab{}.
\newblock \showarticletitle{Toward trustworthy AI development: mechanisms for
  supporting verifiable claims}.
\newblock \bibinfo{journal}{\emph{arXiv preprint arXiv:2004.07213}}
  (\bibinfo{year}{2020}).
\newblock


\bibitem[Cetinic and She(2022)]%
        {10.1145/3475799}
\bibfield{author}{\bibinfo{person}{Eva Cetinic} {and} \bibinfo{person}{James
  She}.} \bibinfo{year}{2022}\natexlab{}.
\newblock \showarticletitle{Understanding and Creating Art with AI: Review and
  Outlook}.
\newblock \bibinfo{journal}{\emph{ACM Trans. Multimedia Comput. Commun. Appl.}}
  \bibinfo{volume}{18}, \bibinfo{number}{2}, Article \bibinfo{articleno}{66}
  (\bibinfo{date}{feb} \bibinfo{year}{2022}), \bibinfo{numpages}{22}~pages.
\newblock
\showISSN{1551-6857}
\urldef\tempurl%
\url{https://doi.org/10.1145/3475799}
\showDOI{\tempurl}


\bibitem[Chayka(2023)]%
        {chayka2023is}
\bibfield{author}{\bibinfo{person}{Kyle Chayka}.}
  \bibinfo{year}{2023}\natexlab{}.
\newblock \bibinfo{booktitle}{\emph{Is A.I. Art Stealing from Artists?}}
\newblock
\urldef\tempurl%
\url{https://www.newyorker.com/culture/infinite-scroll/is-ai-art-stealing-from-artists}
\showURL{%
\tempurl}


\bibitem[Coeckelbergh(2023)]%
        {coeckelbergh2023work}
\bibfield{author}{\bibinfo{person}{Mark Coeckelbergh}.}
  \bibinfo{year}{2023}\natexlab{}.
\newblock \showarticletitle{The Work of Art in the Age of AI Image Generation:
  Aesthetics and Human-Technology Relations as Process and Performance}.
\newblock \bibinfo{journal}{\emph{Journal of Human-Technology Relations}}
  \bibinfo{volume}{1} (\bibinfo{year}{2023}).
\newblock


\bibitem[Cohn(2018)]%
        {gabe2018up}
\bibfield{author}{\bibinfo{person}{Gabe Cohn}.}
  \bibinfo{year}{2018}\natexlab{}.
\newblock \showarticletitle{Up for Bid, AI Art Signed ‘Algorithm’}.
\newblock \bibinfo{journal}{\emph{The New York Times}} \bibinfo{volume}{2},
  \bibinfo{number}{October} (\bibinfo{year}{2018}).
\newblock
\urldef\tempurl%
\url{https://www.nytimes.com/2018/10/22/arts/design/christies-art-artificial-intelligence-obvious.html}
\showURL{%
\tempurl}


\bibitem[Cohn(2023)]%
        {stackpole2023why}
\bibfield{author}{\bibinfo{person}{Gabe Cohn}.}
  \bibinfo{year}{2023}\natexlab{}.
\newblock \bibinfo{booktitle}{\emph{Why generative AI needs a creative human
  touch}}.
\newblock
\urldef\tempurl%
\url{https://mitsloan.mit.edu/ideas-made-to-matter/why-generative-ai-needs-a-creative-human-touch}
\showURL{%
\tempurl}


\bibitem[Commission et~al\mbox{.}(2022)]%
        {doi/10.2759/570559}
\bibfield{author}{\bibinfo{person}{European Commission},
  \bibinfo{person}{Content Directorate-General~for Communications~Networks},
  {and} \bibinfo{person}{Technology}.} \bibinfo{year}{2022}\natexlab{}.
\newblock \bibinfo{booktitle}{\emph{Study on copyright and new technologies :
  copyright data management and artificial intelligence}}.
\newblock \bibinfo{publisher}{Publications Office of the European Union}.
\newblock
\urldef\tempurl%
\url{https://doi.org/doi/10.2759/570559}
\showDOI{\tempurl}


\bibitem[Commission(2023)]%
        {FTC2023generative}
\bibfield{author}{\bibinfo{person}{Federal~Trade Commission}.}
  \bibinfo{year}{2023}\natexlab{}.
\newblock \bibinfo{booktitle}{\emph{Generative AI Raises Competition
  Concerns}}.
\newblock
\urldef\tempurl%
\url{https://www.ftc.gov/policy/advocacy-research/tech-at-ftc/2023/06/generative-ai-raises-competition-concerns}
\showURL{%
\tempurl}


\bibitem[Cui et~al\mbox{.}(2022)]%
        {10.1145/3503161.3548433}
\bibfield{author}{\bibinfo{person}{Shenglan Cui}, \bibinfo{person}{Fang Liu},
  \bibinfo{person}{Tongqing Zhou}, {and} \bibinfo{person}{Mohan Zhang}.}
  \bibinfo{year}{2022}\natexlab{}.
\newblock \showarticletitle{Understanding and Identifying Artwork Plagiarism
  with the Wisdom of Designers: A Case Study on Poster Artworks}. In
  \bibinfo{booktitle}{\emph{Proceedings of the 30th ACM International
  Conference on Multimedia}} (Lisboa, Portugal) \emph{(\bibinfo{series}{MM
  '22})}. \bibinfo{publisher}{Association for Computing Machinery},
  \bibinfo{address}{New York, NY, USA}, \bibinfo{pages}{1117–1127}.
\newblock
\showISBNx{9781450392037}
\urldef\tempurl%
\url{https://doi.org/10.1145/3503161.3548433}
\showDOI{\tempurl}


\bibitem[DALTON(2023)]%
        {strike}
\bibfield{author}{\bibinfo{person}{ANDREW DALTON}.}
  \bibinfo{year}{2023}\natexlab{}.
\newblock \showarticletitle{AI is the wild card in Hollywood’s strikes.
  Here’s an explanation of its unsettling role}.
\newblock \bibinfo{journal}{\emph{The Associated Press}}
  (\bibinfo{year}{2023}).
\newblock
\urldef\tempurl%
\url{https://apnews.com/article/artificial-intelligence-hollywood-strikes-explained-writers-actors-e872bd63ab52c3ea9f7d6e825240a202}
\showURL{%
\tempurl}


\bibitem[Danezis et~al\mbox{.}(2015)]%
        {danezis2015privacy}
\bibfield{author}{\bibinfo{person}{George Danezis}, \bibinfo{person}{Josep
  Domingo-Ferrer}, \bibinfo{person}{Marit Hansen}, \bibinfo{person}{Jaap-Henk
  Hoepman}, \bibinfo{person}{Daniel~Le Metayer}, \bibinfo{person}{Rodica
  Tirtea}, {and} \bibinfo{person}{Stefan Schiffner}.}
  \bibinfo{year}{2015}\natexlab{}.
\newblock \showarticletitle{Privacy and data protection by design-from policy
  to engineering}.
\newblock \bibinfo{journal}{\emph{arXiv preprint arXiv:1501.03726}}
  (\bibinfo{year}{2015}).
\newblock


\bibitem[Denning(2023)]%
        {10.1145/3592981}
\bibfield{author}{\bibinfo{person}{Peter~J. Denning}.}
  \bibinfo{year}{2023}\natexlab{}.
\newblock \showarticletitle{Can Generative AI Bots Be Trusted?}
\newblock \bibinfo{journal}{\emph{Commun. ACM}} \bibinfo{volume}{66},
  \bibinfo{number}{6} (\bibinfo{date}{may} \bibinfo{year}{2023}),
  \bibinfo{pages}{24–27}.
\newblock
\showISSN{0001-0782}
\urldef\tempurl%
\url{https://doi.org/10.1145/3592981}
\showDOI{\tempurl}


\bibitem[Devlin et~al\mbox{.}(2018)]%
        {devlin2018bert}
\bibfield{author}{\bibinfo{person}{Jacob Devlin}, \bibinfo{person}{Ming-Wei
  Chang}, \bibinfo{person}{Kenton Lee}, {and} \bibinfo{person}{Kristina
  Toutanova}.} \bibinfo{year}{2018}\natexlab{}.
\newblock \showarticletitle{Bert: Pre-training of deep bidirectional
  transformers for language understanding}.
\newblock \bibinfo{journal}{\emph{arXiv preprint arXiv:1810.04805}}
  (\bibinfo{year}{2018}).
\newblock


\bibitem[Dhariwal and Nichol(2021)]%
        {NEURIPS2021_49ad23d1}
\bibfield{author}{\bibinfo{person}{Prafulla Dhariwal} {and}
  \bibinfo{person}{Alexander Nichol}.} \bibinfo{year}{2021}\natexlab{}.
\newblock \showarticletitle{Diffusion Models Beat GANs on Image Synthesis}. In
  \bibinfo{booktitle}{\emph{Advances in Neural Information Processing
  Systems}}, \bibfield{editor}{\bibinfo{person}{M.~Ranzato},
  \bibinfo{person}{A.~Beygelzimer}, \bibinfo{person}{Y.~Dauphin},
  \bibinfo{person}{P.S. Liang}, {and} \bibinfo{person}{J.~Wortman Vaughan}}
  (Eds.), Vol.~\bibinfo{volume}{34}. \bibinfo{publisher}{Curran Associates,
  Inc.}, \bibinfo{pages}{8780--8794}.
\newblock
\urldef\tempurl%
\url{https://proceedings.neurips.cc/paper_files/paper/2021/file/49ad23d1ec9fa4bd8d77d02681df5cfa-Paper.pdf}
\showURL{%
\tempurl}


\bibitem[Dien(2023)]%
        {DIEN2023108621}
\bibfield{author}{\bibinfo{person}{Joseph Dien}.}
  \bibinfo{year}{2023}\natexlab{}.
\newblock \showarticletitle{Editorial: Generative artificial intelligence as a
  plagiarism problem}.
\newblock \bibinfo{journal}{\emph{Biological Psychology}}
  \bibinfo{volume}{181} (\bibinfo{year}{2023}), \bibinfo{pages}{108621}.
\newblock
\showISSN{0301-0511}
\urldef\tempurl%
\url{https://doi.org/10.1016/j.biopsycho.2023.108621}
\showDOI{\tempurl}


\bibitem[Elgammal et~al\mbox{.}(2018)]%
        {elgammal2018picasso}
\bibfield{author}{\bibinfo{person}{Ahmed Elgammal}, \bibinfo{person}{Yan Kang},
  {and} \bibinfo{person}{Milko Den~Leeuw}.} \bibinfo{year}{2018}\natexlab{}.
\newblock \showarticletitle{Picasso, matisse, or a fake? Automated analysis of
  drawings at the stroke level for attribution and authentication}. In
  \bibinfo{booktitle}{\emph{Proceedings of the AAAI Conference on Artificial
  Intelligence}}, Vol.~\bibinfo{volume}{32}.
\newblock


\bibitem[Elkins and Chun(2020)]%
        {elkins2020can}
\bibfield{author}{\bibinfo{person}{Katherine Elkins} {and} \bibinfo{person}{Jon
  Chun}.} \bibinfo{year}{2020}\natexlab{}.
\newblock \showarticletitle{Can GPT-3 pass a writer’s Turing test?}
\newblock \bibinfo{journal}{\emph{Journal of Cultural Analytics}}
  \bibinfo{volume}{5}, \bibinfo{number}{2} (\bibinfo{year}{2020}).
\newblock


\bibitem[Epstein et~al\mbox{.}(2023)]%
        {doi:10.1126/science.adh4451}
\bibfield{author}{\bibinfo{person}{Ziv Epstein}, \bibinfo{person}{Aaron
  Hertzmann}, \bibinfo{person}{the Investigators~of Human~Creativity},
  \bibinfo{person}{Memo Akten}, \bibinfo{person}{Hany Farid},
  \bibinfo{person}{Jessica Fjeld}, \bibinfo{person}{Morgan~R. Frank},
  \bibinfo{person}{Matthew Groh}, \bibinfo{person}{Laura Herman},
  \bibinfo{person}{Neil Leach}, \bibinfo{person}{Robert Mahari},
  \bibinfo{person}{Alex~“Sandy” Pentland}, \bibinfo{person}{Olga
  Russakovsky}, \bibinfo{person}{Hope Schroeder}, {and} \bibinfo{person}{Amy
  Smith}.} \bibinfo{year}{2023}\natexlab{}.
\newblock \showarticletitle{Art and the science of generative AI}.
\newblock \bibinfo{journal}{\emph{Science}} \bibinfo{volume}{380},
  \bibinfo{number}{6650} (\bibinfo{year}{2023}), \bibinfo{pages}{1110--1111}.
\newblock
\urldef\tempurl%
\url{https://doi.org/10.1126/science.adh4451}
\showDOI{\tempurl}
\showeprint{https://www.science.org/doi/pdf/10.1126/science.adh4451}


\bibitem[Flick and Worrall(2022)]%
        {flick2022ethics}
\bibfield{author}{\bibinfo{person}{Catherine Flick} {and} \bibinfo{person}{Kyle
  Worrall}.} \bibinfo{year}{2022}\natexlab{}.
\newblock \showarticletitle{The ethics of creative AI}.
\newblock In \bibinfo{booktitle}{\emph{The Language of Creative AI: Practices,
  Aesthetics and Structures}}. \bibinfo{publisher}{Springer},
  \bibinfo{pages}{73--91}.
\newblock


\bibitem[Franceschelli and Musolesi(2023)]%
        {franceschelli2023creativity}
\bibfield{author}{\bibinfo{person}{Giorgio Franceschelli} {and}
  \bibinfo{person}{Mirco Musolesi}.} \bibinfo{year}{2023}\natexlab{}.
\newblock \showarticletitle{On the creativity of large language models}.
\newblock \bibinfo{journal}{\emph{arXiv preprint arXiv:2304.00008}}
  (\bibinfo{year}{2023}).
\newblock


\bibitem[Francke and Bennett(2019)]%
        {francke2019potential}
\bibfield{author}{\bibinfo{person}{Errol Francke} {and}
  \bibinfo{person}{Alexander Bennett}.} \bibinfo{year}{2019}\natexlab{}.
\newblock \showarticletitle{The potential influence of artificial intelligence
  on plagiarism: A higher education perspective}. In
  \bibinfo{booktitle}{\emph{European Conference on the Impact of Artificial
  Intelligence and Robotics (ECIAIR 2019)}}. \bibinfo{pages}{131--140}.
\newblock


\bibitem[Frye(2022)]%
        {frye2022should}
\bibfield{author}{\bibinfo{person}{Brian~L Frye}.}
  \bibinfo{year}{2022}\natexlab{}.
\newblock \showarticletitle{Should using an AI text generator to produce
  academic writing be plagiarism?}
\newblock \bibinfo{journal}{\emph{Fordham Intellectual Property, Media \&
  Entertainment Law Journal, Forthcoming}} (\bibinfo{year}{2022}).
\newblock


\bibitem[Gao et~al\mbox{.}(2020)]%
        {gao2020painting}
\bibfield{author}{\bibinfo{person}{Xiang Gao}, \bibinfo{person}{Yingjie Tian},
  {and} \bibinfo{person}{Zhiquan Qi}.} \bibinfo{year}{2020}\natexlab{}.
\newblock \showarticletitle{RPD-GAN: Learning to Draw Realistic Paintings With
  Generative Adversarial Network}.
\newblock \bibinfo{journal}{\emph{IEEE Transactions on Image Processing}}
  \bibinfo{volume}{29} (\bibinfo{year}{2020}), \bibinfo{pages}{8706--8720}.
\newblock
\urldef\tempurl%
\url{https://doi.org/10.1109/TIP.2020.3018856}
\showDOI{\tempurl}


\bibitem[Ghosh and Fossas(2022)]%
        {ghosh2022can}
\bibfield{author}{\bibinfo{person}{Avijit Ghosh} {and}
  \bibinfo{person}{Genoveva Fossas}.} \bibinfo{year}{2022}\natexlab{}.
\newblock \showarticletitle{Can there be art without an artist?}
\newblock \bibinfo{journal}{\emph{arXiv preprint arXiv:2209.07667}}
  (\bibinfo{year}{2022}).
\newblock


\bibitem[Glaser et~al\mbox{.}(1968)]%
        {glaser1968discovery}
\bibfield{author}{\bibinfo{person}{Barney~G Glaser}, \bibinfo{person}{Anselm~L
  Strauss}, {and} \bibinfo{person}{Elizabeth Strutzel}.}
  \bibinfo{year}{1968}\natexlab{}.
\newblock \showarticletitle{The discovery of grounded theory; strategies for
  qualitative research}.
\newblock \bibinfo{journal}{\emph{Nursing research}} \bibinfo{volume}{17},
  \bibinfo{number}{4} (\bibinfo{year}{1968}), \bibinfo{pages}{364}.
\newblock


\bibitem[Goodfellow et~al\mbox{.}(2014)]%
        {goodfellow2014generative}
\bibfield{author}{\bibinfo{person}{Ian Goodfellow}, \bibinfo{person}{Jean
  Pouget-Abadie}, \bibinfo{person}{Mehdi Mirza}, \bibinfo{person}{Bing Xu},
  \bibinfo{person}{David Warde-Farley}, \bibinfo{person}{Sherjil Ozair},
  \bibinfo{person}{Aaron Courville}, {and} \bibinfo{person}{Yoshua Bengio}.}
  \bibinfo{year}{2014}\natexlab{}.
\newblock \showarticletitle{Generative adversarial nets}.
\newblock \bibinfo{journal}{\emph{Advances in neural information processing
  systems}}  \bibinfo{volume}{27} (\bibinfo{year}{2014}).
\newblock


\bibitem[Haase and Hanel(2023)]%
        {haase2023artificial}
\bibfield{author}{\bibinfo{person}{Jennifer Haase} {and} \bibinfo{person}{Paul
  H.~P. Hanel}.} \bibinfo{year}{2023}\natexlab{}.
\newblock \showarticletitle{Artificial muses: Generative Artificial
  Intelligence Chatbots Have Risen to Human-Level Creativity}.
\newblock  (\bibinfo{year}{2023}).
\newblock
\showeprint[arxiv]{2303.12003}~[cs.AI]


\bibitem[Hacker et~al\mbox{.}(2023)]%
        {10.1145/3593013.3594067}
\bibfield{author}{\bibinfo{person}{Philipp Hacker}, \bibinfo{person}{Andreas
  Engel}, {and} \bibinfo{person}{Marco Mauer}.}
  \bibinfo{year}{2023}\natexlab{}.
\newblock \showarticletitle{Regulating ChatGPT and Other Large Generative AI
  Models}. In \bibinfo{booktitle}{\emph{Proceedings of the 2023 ACM Conference
  on Fairness, Accountability, and Transparency}} (Chicago, IL, USA)
  \emph{(\bibinfo{series}{FAccT '23})}. \bibinfo{publisher}{Association for
  Computing Machinery}, \bibinfo{address}{New York, NY, USA},
  \bibinfo{pages}{1112–1123}.
\newblock
\showISBNx{9798400701924}
\urldef\tempurl%
\url{https://doi.org/10.1145/3593013.3594067}
\showDOI{\tempurl}


\bibitem[Hamburger(1973)]%
        {hamburger1973n}
\bibfield{author}{\bibinfo{person}{Henry Hamburger}.}
  \bibinfo{year}{1973}\natexlab{}.
\newblock \showarticletitle{N-person prisoner's dilemma}.
\newblock \bibinfo{journal}{\emph{Journal of Mathematical Sociology}}
  \bibinfo{volume}{3}, \bibinfo{number}{1} (\bibinfo{year}{1973}),
  \bibinfo{pages}{27--48}.
\newblock


\bibitem[Hardt et~al\mbox{.}(2016)]%
        {NIPS2016_9d268236}
\bibfield{author}{\bibinfo{person}{Moritz Hardt}, \bibinfo{person}{Eric Price},
  \bibinfo{person}{Eric Price}, {and} \bibinfo{person}{Nati Srebro}.}
  \bibinfo{year}{2016}\natexlab{}.
\newblock \showarticletitle{Equality of Opportunity in Supervised Learning}. In
  \bibinfo{booktitle}{\emph{Advances in Neural Information Processing
  Systems}}, \bibfield{editor}{\bibinfo{person}{D.~Lee},
  \bibinfo{person}{M.~Sugiyama}, \bibinfo{person}{U.~Luxburg},
  \bibinfo{person}{I.~Guyon}, {and} \bibinfo{person}{R.~Garnett}} (Eds.),
  Vol.~\bibinfo{volume}{29}. \bibinfo{publisher}{Curran Associates, Inc.}
\newblock
\urldef\tempurl%
\url{https://proceedings.neurips.cc/paper_files/paper/2016/file/9d2682367c3935defcb1f9e247a97c0d-Paper.pdf}
\showURL{%
\tempurl}


\bibitem[Hedges(2023)]%
        {hedges2023artificial}
\bibfield{author}{\bibinfo{person}{Keith~E Hedges}.}
  \bibinfo{year}{2023}\natexlab{}.
\newblock \showarticletitle{Artificial Intelligence (AI) Art Generators in the
  Architectural Design Curricula}. In \bibinfo{booktitle}{\emph{2023 ASEE
  Annual Conference \& Exposition}}.
\newblock


\bibitem[Heider(1988)]%
        {heider1988rashomon}
\bibfield{author}{\bibinfo{person}{Karl~G Heider}.}
  \bibinfo{year}{1988}\natexlab{}.
\newblock \showarticletitle{The Rashomon effect: When ethnographers disagree}.
\newblock \bibinfo{journal}{\emph{American Anthropologist}}
  \bibinfo{volume}{90}, \bibinfo{number}{1} (\bibinfo{year}{1988}),
  \bibinfo{pages}{73--81}.
\newblock


\bibitem[Hertzmann(2022)]%
        {hertzmann2022toward}
\bibfield{author}{\bibinfo{person}{Aaron Hertzmann}.}
  \bibinfo{year}{2022}\natexlab{}.
\newblock \showarticletitle{Toward modeling creative processes for algorithmic
  painting}.
\newblock \bibinfo{journal}{\emph{arXiv preprint arXiv:2205.01605}}
  (\bibinfo{year}{2022}).
\newblock


\bibitem[Ho et~al\mbox{.}(2020)]%
        {ho2020denoising}
\bibfield{author}{\bibinfo{person}{Jonathan Ho}, \bibinfo{person}{Ajay Jain},
  {and} \bibinfo{person}{Pieter Abbeel}.} \bibinfo{year}{2020}\natexlab{}.
\newblock \bibinfo{title}{Denoising Diffusion Probabilistic Models}.
\newblock
\newblock
\showeprint[arxiv]{2006.11239}~[cs.LG]


\bibitem[Hutson and Harper-Nichols(2023)]%
        {hutson2023generative}
\bibfield{author}{\bibinfo{person}{James Hutson} {and} \bibinfo{person}{Morgan
  Harper-Nichols}.} \bibinfo{year}{2023}\natexlab{}.
\newblock \showarticletitle{Generative AI and Algorithmic Art: Disrupting the
  Framing of Meaning and Rethinking the Subject-Object Dilemma}.
\newblock \bibinfo{journal}{\emph{Global Journal of Computer Science and
  Technology: D}} \bibinfo{volume}{23}, \bibinfo{number}{1}
  (\bibinfo{year}{2023}).
\newblock


\bibitem[Hutson and Lang(2023)]%
        {hutson2023content}
\bibfield{author}{\bibinfo{person}{James Hutson} {and} \bibinfo{person}{Martin
  Lang}.} \bibinfo{year}{2023}\natexlab{}.
\newblock \showarticletitle{Content creation or interpolation: AI generative
  digital art in the classroom}.
\newblock \bibinfo{journal}{\emph{Metaverse}} \bibinfo{volume}{4},
  \bibinfo{number}{1} (\bibinfo{year}{2023}).
\newblock


\bibitem[Kamran et~al\mbox{.}(2021)]%
        {kamran2021rv}
\bibfield{author}{\bibinfo{person}{Sharif~Amit Kamran},
  \bibinfo{person}{Khondker~Fariha Hossain}, \bibinfo{person}{Alireza
  Tavakkoli}, \bibinfo{person}{Stewart~Lee Zuckerbrod},
  \bibinfo{person}{Kenton~M Sanders}, {and} \bibinfo{person}{Salah~A Baker}.}
  \bibinfo{year}{2021}\natexlab{}.
\newblock \showarticletitle{RV-GAN: Segmenting retinal vascular structure in
  fundus photographs using a novel multi-scale generative adversarial network}.
  In \bibinfo{booktitle}{\emph{Medical Image Computing and Computer Assisted
  Intervention--MICCAI 2021: 24th International Conference, Strasbourg, France,
  September 27--October 1, 2021, Proceedings, Part VIII 24}}. Springer,
  \bibinfo{pages}{34--44}.
\newblock


\bibitem[Karras et~al\mbox{.}(2017)]%
        {karras2017progressive}
\bibfield{author}{\bibinfo{person}{Tero Karras}, \bibinfo{person}{Timo Aila},
  \bibinfo{person}{Samuli Laine}, {and} \bibinfo{person}{Jaakko Lehtinen}.}
  \bibinfo{year}{2017}\natexlab{}.
\newblock \showarticletitle{Progressive growing of gans for improved quality,
  stability, and variation}.
\newblock \bibinfo{journal}{\emph{arXiv preprint arXiv:1710.10196}}
  (\bibinfo{year}{2017}).
\newblock


\bibitem[Karras et~al\mbox{.}(2019)]%
        {karras2019style}
\bibfield{author}{\bibinfo{person}{Tero Karras}, \bibinfo{person}{Samuli
  Laine}, {and} \bibinfo{person}{Timo Aila}.} \bibinfo{year}{2019}\natexlab{}.
\newblock \showarticletitle{A style-based generator architecture for generative
  adversarial networks}. In \bibinfo{booktitle}{\emph{Proceedings of the
  IEEE/CVF conference on computer vision and pattern recognition}}.
  \bibinfo{pages}{4401--4410}.
\newblock


\bibitem[Keskar et~al\mbox{.}(2019)]%
        {keskar2019ctrl}
\bibfield{author}{\bibinfo{person}{Nitish~Shirish Keskar},
  \bibinfo{person}{Bryan McCann}, \bibinfo{person}{Lav~R Varshney},
  \bibinfo{person}{Caiming Xiong}, {and} \bibinfo{person}{Richard Socher}.}
  \bibinfo{year}{2019}\natexlab{}.
\newblock \showarticletitle{Ctrl: A conditional transformer language model for
  controllable generation}.
\newblock \bibinfo{journal}{\emph{arXiv preprint arXiv:1909.05858}}
  (\bibinfo{year}{2019}).
\newblock


\bibitem[Kulkarni et~al\mbox{.}(2019)]%
        {kulkarni2019survey}
\bibfield{author}{\bibinfo{person}{Rajat Kulkarni}, \bibinfo{person}{Rutik
  Gaikwad}, \bibinfo{person}{Rudraksh Sugandhi}, \bibinfo{person}{Pranjali
  Kulkarni}, {and} \bibinfo{person}{Shivraj Kone}.}
  \bibinfo{year}{2019}\natexlab{}.
\newblock \showarticletitle{Survey on deep learning in music using GAN}.
\newblock \bibinfo{journal}{\emph{International Journal of Engineering Research
  \& Technology}} \bibinfo{volume}{8}, \bibinfo{number}{9}
  (\bibinfo{year}{2019}), \bibinfo{pages}{646--648}.
\newblock


\bibitem[Li et~al\mbox{.}(2020b)]%
        {li2020attribute}
\bibfield{author}{\bibinfo{person}{Jianan Li}, \bibinfo{person}{Jimei Yang},
  \bibinfo{person}{Jianming Zhang}, \bibinfo{person}{Chang Liu},
  \bibinfo{person}{Christina Wang}, {and} \bibinfo{person}{Tingfa Xu}.}
  \bibinfo{year}{2020}\natexlab{b}.
\newblock \showarticletitle{Attribute-conditioned layout gan for automatic
  graphic design}.
\newblock \bibinfo{journal}{\emph{IEEE Transactions on Visualization and
  Computer Graphics}} \bibinfo{volume}{27}, \bibinfo{number}{10}
  (\bibinfo{year}{2020}), \bibinfo{pages}{4039--4048}.
\newblock


\bibitem[Li and Sung(2021)]%
        {li2021inco}
\bibfield{author}{\bibinfo{person}{Shuyu Li} {and} \bibinfo{person}{Yunsick
  Sung}.} \bibinfo{year}{2021}\natexlab{}.
\newblock \showarticletitle{INCO-GAN: variable-length music generation method
  based on inception model-based conditional GAN}.
\newblock \bibinfo{journal}{\emph{Mathematics}} \bibinfo{volume}{9},
  \bibinfo{number}{4} (\bibinfo{year}{2021}), \bibinfo{pages}{387}.
\newblock


\bibitem[Li et~al\mbox{.}(2019)]%
        {li2019storygan}
\bibfield{author}{\bibinfo{person}{Yitong Li}, \bibinfo{person}{Zhe Gan},
  \bibinfo{person}{Yelong Shen}, \bibinfo{person}{Jingjing Liu},
  \bibinfo{person}{Yu Cheng}, \bibinfo{person}{Yuexin Wu},
  \bibinfo{person}{Lawrence Carin}, \bibinfo{person}{David Carlson}, {and}
  \bibinfo{person}{Jianfeng Gao}.} \bibinfo{year}{2019}\natexlab{}.
\newblock \showarticletitle{Storygan: A sequential conditional gan for story
  visualization}. In \bibinfo{booktitle}{\emph{Proceedings of the IEEE/CVF
  Conference on Computer Vision and Pattern Recognition}}.
  \bibinfo{pages}{6329--6338}.
\newblock


\bibitem[Li et~al\mbox{.}(2020a)]%
        {li2020research}
\bibfield{author}{\bibinfo{person}{Yueen Li}, \bibinfo{person}{Jin Gu}, {and}
  \bibinfo{person}{Liyang Wang}.} \bibinfo{year}{2020}\natexlab{a}.
\newblock \showarticletitle{Research on artificial intelligence ethics in the
  field of art design}. In \bibinfo{booktitle}{\emph{Journal of Physics:
  Conference Series}}, Vol.~\bibinfo{volume}{1673}. IOP Publishing,
  \bibinfo{pages}{012052}.
\newblock


\bibitem[Liao(2020)]%
        {liao2020ethics}
\bibfield{author}{\bibinfo{person}{S.M. Liao}.}
  \bibinfo{year}{2020}\natexlab{}.
\newblock \bibinfo{booktitle}{\emph{Ethics of Artificial Intelligence}}.
\newblock \bibinfo{publisher}{Oxford University Press}.
\newblock
\showISBNx{9780190905057}
\showLCCN{2020004474}
\urldef\tempurl%
\url{https://books.google.com/books?id=1yT3DwAAQBAJ}
\showURL{%
\tempurl}


\bibitem[Liao et~al\mbox{.}(2022)]%
        {Liao_2022_CVPR}
\bibfield{author}{\bibinfo{person}{Wentong Liao}, \bibinfo{person}{Kai Hu},
  \bibinfo{person}{Michael~Ying Yang}, {and} \bibinfo{person}{Bodo Rosenhahn}.}
  \bibinfo{year}{2022}\natexlab{}.
\newblock \showarticletitle{Text to Image Generation With Semantic-Spatial
  Aware GAN}. In \bibinfo{booktitle}{\emph{Proceedings of the IEEE/CVF
  Conference on Computer Vision and Pattern Recognition (CVPR)}}.
  \bibinfo{pages}{18187--18196}.
\newblock


\bibitem[Liu(2023)]%
        {liu2023arguments}
\bibfield{author}{\bibinfo{person}{Bai Liu}.} \bibinfo{year}{2023}\natexlab{}.
\newblock \showarticletitle{Arguments for the Rise of Artificial Intelligence
  Art: Does AI Art Have Creativity, Motivation, Self-awareness and Emotion?}
\newblock \bibinfo{journal}{\emph{Arte}}  \bibinfo{volume}{Avance en línea}
  (\bibinfo{date}{04} \bibinfo{year}{2023}), \bibinfo{pages}{1--11}.
\newblock
\urldef\tempurl%
\url{https://doi.org/10.5209/aris.83808}
\showDOI{\tempurl}


\bibitem[Markuckas(2022)]%
        {markuckas2022question}
\bibfield{author}{\bibinfo{person}{Marius Markuckas}.}
  \bibinfo{year}{2022}\natexlab{}.
\newblock \bibinfo{title}{On the Question of the Possibility to Replace the
  Human with Technology in the Creative Process}.
\newblock
\newblock


\bibitem[Mehrabi et~al\mbox{.}(2021)]%
        {mehrabi2021survey}
\bibfield{author}{\bibinfo{person}{Ninareh Mehrabi}, \bibinfo{person}{Fred
  Morstatter}, \bibinfo{person}{Nripsuta Saxena}, \bibinfo{person}{Kristina
  Lerman}, {and} \bibinfo{person}{Aram Galstyan}.}
  \bibinfo{year}{2021}\natexlab{}.
\newblock \showarticletitle{A survey on bias and fairness in machine learning}.
\newblock \bibinfo{journal}{\emph{ACM computing surveys (CSUR)}}
  \bibinfo{volume}{54}, \bibinfo{number}{6} (\bibinfo{year}{2021}),
  \bibinfo{pages}{1--35}.
\newblock


\bibitem[Mikalonyte and Kneer({[n.\,d.]})]%
        {mikalonytefolk}
\bibfield{author}{\bibinfo{person}{Elze~Sigute Mikalonyte} {and}
  \bibinfo{person}{Markus Kneer}.} \bibinfo{year}{[n.\,d.]}\natexlab{}.
\newblock \showarticletitle{The Folk Concept of Art}.
\newblock  (\bibinfo{year}{[n.\,d.]}).
\newblock


\bibitem[Mikalonyt{\.e} and Kneer(2022)]%
        {mikalonyte2022can}
\bibfield{author}{\bibinfo{person}{Elz{\.e}~Sigut{\.e} Mikalonyt{\.e}} {and}
  \bibinfo{person}{Markus Kneer}.} \bibinfo{year}{2022}\natexlab{}.
\newblock \showarticletitle{Can artificial intelligence make art?: Folk
  intuitions as to whether AI-driven robots can be viewed as artists and
  produce art}.
\newblock \bibinfo{journal}{\emph{ACM Transactions on Human-Robot Interaction
  (THRI)}} \bibinfo{volume}{11}, \bibinfo{number}{4} (\bibinfo{year}{2022}),
  \bibinfo{pages}{1--19}.
\newblock


\bibitem[Mikalonyt{\.e} and Kneer(2023)]%
        {mikalonyte2023art}
\bibfield{author}{\bibinfo{person}{Elz{\.e}~Sigut{\.e} Mikalonyt{\.e}} {and}
  \bibinfo{person}{Markus Kneer}.} \bibinfo{year}{2023}\natexlab{}.
\newblock \showarticletitle{What Is Art? The Role of Intention, Beauty, and
  Institutional Recognition}. In \bibinfo{booktitle}{\emph{Proceedings of the
  Annual Meeting of the Cognitive Science Society}}, Vol.~\bibinfo{volume}{45}.
\newblock


\bibitem[Mirza and Osindero(2014)]%
        {mirza2014conditional}
\bibfield{author}{\bibinfo{person}{Mehdi Mirza} {and} \bibinfo{person}{Simon
  Osindero}.} \bibinfo{year}{2014}\natexlab{}.
\newblock \showarticletitle{Conditional generative adversarial nets}.
\newblock \bibinfo{journal}{\emph{arXiv preprint arXiv:1411.1784}}
  (\bibinfo{year}{2014}).
\newblock


\bibitem[Newton and Dhole(2023)]%
        {newton2023ai}
\bibfield{author}{\bibinfo{person}{Alexis Newton} {and}
  \bibinfo{person}{Kaustubh Dhole}.} \bibinfo{year}{2023}\natexlab{}.
\newblock \showarticletitle{Is AI Art Another Industrial Revolution in the
  Making?}
\newblock \bibinfo{journal}{\emph{arXiv preprint arXiv:2301.05133}}
  (\bibinfo{year}{2023}).
\newblock


\bibitem[Nichol et~al\mbox{.}(2022)]%
        {nichol2022glide}
\bibfield{author}{\bibinfo{person}{Alex Nichol}, \bibinfo{person}{Prafulla
  Dhariwal}, \bibinfo{person}{Aditya Ramesh}, \bibinfo{person}{Pranav Shyam},
  \bibinfo{person}{Pamela Mishkin}, \bibinfo{person}{Bob McGrew},
  \bibinfo{person}{Ilya Sutskever}, {and} \bibinfo{person}{Mark Chen}.}
  \bibinfo{year}{2022}\natexlab{}.
\newblock \bibinfo{title}{GLIDE: Towards Photorealistic Image Generation and
  Editing with Text-Guided Diffusion Models}.
\newblock
\newblock
\showeprint[arxiv]{2112.10741}~[cs.CV]


\bibitem[Noci(2023)]%
        {noci2023merging}
\bibfield{author}{\bibinfo{person}{Javier~D{\'\i}az Noci}.}
  \bibinfo{year}{2023}\natexlab{}.
\newblock \showarticletitle{Merging or plagiarizing? The role of originality
  and derivative works in AI-aimed news production}.
\newblock \bibinfo{journal}{\emph{Hipertext. net}} \bibinfo{number}{26}
  (\bibinfo{year}{2023}), \bibinfo{pages}{69--76}.
\newblock


\bibitem[Pedregosa and Eleni~Triantafillou(2023)]%
        {machineunlearning}
\bibfield{author}{\bibinfo{person}{Fabian Pedregosa} {and}
  \bibinfo{person}{Google Eleni~Triantafillou, Research~Scientists}.}
  \bibinfo{year}{2023}\natexlab{}.
\newblock \showarticletitle{Announcing the first Machine Unlearning Challenge}.
\newblock  (\bibinfo{year}{2023}).
\newblock
\urldef\tempurl%
\url{https://ai.googleblog.com/2023/06/announcing-first-machine-unlearning.html}
\showURL{%
\tempurl}
\newblock
\shownote{Last accessed on August 2, 2023}.


\bibitem[Peng et~al\mbox{.}(2019)]%
        {peng2019poetry}
\bibfield{author}{\bibinfo{person}{Guan-Fu Peng}, \bibinfo{person}{Yi-Shian
  Yang}, \bibinfo{person}{Chen-Yu Tsai}, {and} \bibinfo{person}{Von-Wun Soo}.}
  \bibinfo{year}{2019}\natexlab{}.
\newblock \showarticletitle{Generate Modern Chinese Poems from News Based on
  Text Style Transfer Using GAN}. In \bibinfo{booktitle}{\emph{2019
  International Conference on Technologies and Applications of Artiﬁcial
  Intelligence (TAAI)}}. \bibinfo{pages}{1--6}.
\newblock
\urldef\tempurl%
\url{https://doi.org/10.1109/TAAI48200.2019.8959907}
\showDOI{\tempurl}


\bibitem[Pente et~al\mbox{.}(2022)]%
        {pente2022artificial}
\bibfield{author}{\bibinfo{person}{Patti Pente}, \bibinfo{person}{Catherine
  Adams}, {and} \bibinfo{person}{Connie Yuen}.}
  \bibinfo{year}{2022}\natexlab{}.
\newblock \showarticletitle{Artificial Intelligence, ethics, and art education
  in a posthuman world}.
\newblock In \bibinfo{booktitle}{\emph{Global Media Arts Education: Mapping
  Global Perspectives of Media Arts in Education}}.
  \bibinfo{publisher}{Springer}, \bibinfo{pages}{197--211}.
\newblock


\bibitem[Radford et~al\mbox{.}(2021)]%
        {radford2021learning}
\bibfield{author}{\bibinfo{person}{Alec Radford}, \bibinfo{person}{Jong~Wook
  Kim}, \bibinfo{person}{Chris Hallacy}, \bibinfo{person}{Aditya Ramesh},
  \bibinfo{person}{Gabriel Goh}, \bibinfo{person}{Sandhini Agarwal},
  \bibinfo{person}{Girish Sastry}, \bibinfo{person}{Amanda Askell},
  \bibinfo{person}{Pamela Mishkin}, \bibinfo{person}{Jack Clark},
  \bibinfo{person}{Gretchen Krueger}, {and} \bibinfo{person}{Ilya Sutskever}.}
  \bibinfo{year}{2021}\natexlab{}.
\newblock \bibinfo{title}{Learning Transferable Visual Models From Natural
  Language Supervision}.
\newblock
\newblock
\showeprint[arxiv]{2103.00020}~[cs.CV]


\bibitem[Radford et~al\mbox{.}(2015)]%
        {radford2015unsupervised}
\bibfield{author}{\bibinfo{person}{Alec Radford}, \bibinfo{person}{Luke Metz},
  {and} \bibinfo{person}{Soumith Chintala}.} \bibinfo{year}{2015}\natexlab{}.
\newblock \showarticletitle{Unsupervised representation learning with deep
  convolutional generative adversarial networks}.
\newblock \bibinfo{journal}{\emph{arXiv preprint arXiv:1511.06434}}
  (\bibinfo{year}{2015}).
\newblock


\bibitem[Radford et~al\mbox{.}(2018)]%
        {radford2018improving}
\bibfield{author}{\bibinfo{person}{Alec Radford}, \bibinfo{person}{Karthik
  Narasimhan}, \bibinfo{person}{Tim Salimans}, \bibinfo{person}{Ilya
  Sutskever}, {et~al\mbox{.}}} \bibinfo{year}{2018}\natexlab{}.
\newblock \showarticletitle{Improving language understanding by generative
  pre-training}.
\newblock  (\bibinfo{year}{2018}).
\newblock


\bibitem[Radford et~al\mbox{.}(2019)]%
        {radford2019language}
\bibfield{author}{\bibinfo{person}{Alec Radford}, \bibinfo{person}{Jeffrey Wu},
  \bibinfo{person}{Rewon Child}, \bibinfo{person}{David Luan},
  \bibinfo{person}{Dario Amodei}, \bibinfo{person}{Ilya Sutskever},
  {et~al\mbox{.}}} \bibinfo{year}{2019}\natexlab{}.
\newblock \showarticletitle{Language models are unsupervised multitask
  learners}.
\newblock \bibinfo{journal}{\emph{OpenAI blog}} \bibinfo{volume}{1},
  \bibinfo{number}{8} (\bibinfo{year}{2019}), \bibinfo{pages}{9}.
\newblock


\bibitem[Raffel et~al\mbox{.}(2020)]%
        {raffel2020exploring}
\bibfield{author}{\bibinfo{person}{Colin Raffel}, \bibinfo{person}{Noam
  Shazeer}, \bibinfo{person}{Adam Roberts}, \bibinfo{person}{Katherine Lee},
  \bibinfo{person}{Sharan Narang}, \bibinfo{person}{Michael Matena},
  \bibinfo{person}{Yanqi Zhou}, \bibinfo{person}{Wei Li}, {and}
  \bibinfo{person}{Peter~J. Liu}.} \bibinfo{year}{2020}\natexlab{}.
\newblock \bibinfo{title}{Exploring the Limits of Transfer Learning with a
  Unified Text-to-Text Transformer}.
\newblock
\newblock
\showeprint[arxiv]{1910.10683}~[cs.LG]


\bibitem[Ragot et~al\mbox{.}(2020)]%
        {10.1145/3334480.3382892}
\bibfield{author}{\bibinfo{person}{Martin Ragot}, \bibinfo{person}{Nicolas
  Martin}, {and} \bibinfo{person}{Salom\'{e} Cojean}.}
  \bibinfo{year}{2020}\natexlab{}.
\newblock \showarticletitle{AI-Generated vs. Human Artworks. A Perception Bias
  Towards Artificial Intelligence?}. In \bibinfo{booktitle}{\emph{Extended
  Abstracts of the 2020 CHI Conference on Human Factors in Computing Systems}}
  (Honolulu, HI, USA) \emph{(\bibinfo{series}{CHI EA '20})}.
  \bibinfo{publisher}{Association for Computing Machinery},
  \bibinfo{address}{New York, NY, USA}, \bibinfo{pages}{1–10}.
\newblock
\showISBNx{9781450368193}
\urldef\tempurl%
\url{https://doi.org/10.1145/3334480.3382892}
\showDOI{\tempurl}


\bibitem[Ramesh et~al\mbox{.}(2022)]%
        {ramesh2022hierarchical}
\bibfield{author}{\bibinfo{person}{Aditya Ramesh}, \bibinfo{person}{Prafulla
  Dhariwal}, \bibinfo{person}{Alex Nichol}, \bibinfo{person}{Casey Chu}, {and}
  \bibinfo{person}{Mark Chen}.} \bibinfo{year}{2022}\natexlab{}.
\newblock \bibinfo{title}{Hierarchical Text-Conditional Image Generation with
  CLIP Latents}.
\newblock
\newblock
\showeprint[arxiv]{2204.06125}~[cs.CV]


\bibitem[Ramesh et~al\mbox{.}(2021)]%
        {ramesh2021zeroshot}
\bibfield{author}{\bibinfo{person}{Aditya Ramesh}, \bibinfo{person}{Mikhail
  Pavlov}, \bibinfo{person}{Gabriel Goh}, \bibinfo{person}{Scott Gray},
  \bibinfo{person}{Chelsea Voss}, \bibinfo{person}{Alec Radford},
  \bibinfo{person}{Mark Chen}, {and} \bibinfo{person}{Ilya Sutskever}.}
  \bibinfo{year}{2021}\natexlab{}.
\newblock \bibinfo{title}{Zero-Shot Text-to-Image Generation}.
\newblock
\newblock
\showeprint[arxiv]{2102.12092}~[cs.CV]


\bibitem[Rombach et~al\mbox{.}(2022)]%
        {Rombach_2022_CVPR}
\bibfield{author}{\bibinfo{person}{Robin Rombach}, \bibinfo{person}{Andreas
  Blattmann}, \bibinfo{person}{Dominik Lorenz}, \bibinfo{person}{Patrick
  Esser}, {and} \bibinfo{person}{Bj\"orn Ommer}.}
  \bibinfo{year}{2022}\natexlab{}.
\newblock \showarticletitle{High-Resolution Image Synthesis With Latent
  Diffusion Models}. In \bibinfo{booktitle}{\emph{Proceedings of the IEEE/CVF
  Conference on Computer Vision and Pattern Recognition (CVPR)}}.
  \bibinfo{pages}{10684--10695}.
\newblock


\bibitem[Roose(2022)]%
        {roose2022ai}
\bibfield{author}{\bibinfo{person}{Kevin Roose}.}
  \bibinfo{year}{2022}\natexlab{}.
\newblock \showarticletitle{An AI-generated picture won an art prize. Artists
  aren’t happy}.
\newblock \bibinfo{journal}{\emph{The New York Times}} \bibinfo{volume}{2},
  \bibinfo{number}{September} (\bibinfo{year}{2022}).
\newblock
\urldef\tempurl%
\url{https://www.nytimes.com/2022/09/02/technology/ai-artificial-intelligence-artists.html}
\showURL{%
\tempurl}


\bibitem[Saharia et~al\mbox{.}(2022a)]%
        {saharia2022photorealistic}
\bibfield{author}{\bibinfo{person}{Chitwan Saharia}, \bibinfo{person}{William
  Chan}, \bibinfo{person}{Saurabh Saxena}, \bibinfo{person}{Lala Li},
  \bibinfo{person}{Jay Whang}, \bibinfo{person}{Emily Denton},
  \bibinfo{person}{Seyed Kamyar~Seyed Ghasemipour},
  \bibinfo{person}{Burcu~Karagol Ayan}, \bibinfo{person}{S.~Sara Mahdavi},
  \bibinfo{person}{Rapha~Gontijo Lopes}, \bibinfo{person}{Tim Salimans},
  \bibinfo{person}{Jonathan Ho}, \bibinfo{person}{David~J Fleet}, {and}
  \bibinfo{person}{Mohammad Norouzi}.} \bibinfo{year}{2022}\natexlab{a}.
\newblock \bibinfo{title}{Photorealistic Text-to-Image Diffusion Models with
  Deep Language Understanding}.
\newblock
\newblock
\showeprint[arxiv]{2205.11487}~[cs.CV]


\bibitem[Saharia et~al\mbox{.}(2022b)]%
        {NEURIPS2022_ec795aea}
\bibfield{author}{\bibinfo{person}{Chitwan Saharia}, \bibinfo{person}{William
  Chan}, \bibinfo{person}{Saurabh Saxena}, \bibinfo{person}{Lala Li},
  \bibinfo{person}{Jay Whang}, \bibinfo{person}{Emily~L Denton},
  \bibinfo{person}{Kamyar Ghasemipour}, \bibinfo{person}{Raphael
  Gontijo~Lopes}, \bibinfo{person}{Burcu Karagol~Ayan}, \bibinfo{person}{Tim
  Salimans}, \bibinfo{person}{Jonathan Ho}, \bibinfo{person}{David~J Fleet},
  {and} \bibinfo{person}{Mohammad Norouzi}.} \bibinfo{year}{2022}\natexlab{b}.
\newblock \showarticletitle{Photorealistic Text-to-Image Diffusion Models with
  Deep Language Understanding}. In \bibinfo{booktitle}{\emph{Advances in Neural
  Information Processing Systems}},
  \bibfield{editor}{\bibinfo{person}{S.~Koyejo}, \bibinfo{person}{S.~Mohamed},
  \bibinfo{person}{A.~Agarwal}, \bibinfo{person}{D.~Belgrave},
  \bibinfo{person}{K.~Cho}, {and} \bibinfo{person}{A.~Oh}} (Eds.),
  Vol.~\bibinfo{volume}{35}. \bibinfo{publisher}{Curran Associates, Inc.},
  \bibinfo{pages}{36479--36494}.
\newblock
\urldef\tempurl%
\url{https://proceedings.neurips.cc/paper_files/paper/2022/file/ec795aeadae0b7d230fa35cbaf04c041-Paper-Conference.pdf}
\showURL{%
\tempurl}


\bibitem[Salimans et~al\mbox{.}(2016)]%
        {salimans2016improved}
\bibfield{author}{\bibinfo{person}{Tim Salimans}, \bibinfo{person}{Ian
  Goodfellow}, \bibinfo{person}{Wojciech Zaremba}, \bibinfo{person}{Vicki
  Cheung}, \bibinfo{person}{Alec Radford}, {and} \bibinfo{person}{Xi Chen}.}
  \bibinfo{year}{2016}\natexlab{}.
\newblock \showarticletitle{Improved techniques for training gans}.
\newblock \bibinfo{journal}{\emph{Advances in neural information processing
  systems}}  \bibinfo{volume}{29} (\bibinfo{year}{2016}).
\newblock


\bibitem[Sarkar(2023)]%
        {sarkar2023exploring}
\bibfield{author}{\bibinfo{person}{Advait Sarkar}.}
  \bibinfo{year}{2023}\natexlab{}.
\newblock \showarticletitle{Exploring Perspectives on the Impact of Artificial
  Intelligence on the Creativity of Knowledge Work: Beyond Mechanised
  Plagiarism and Stochastic Parrots}.
\newblock  (\bibinfo{year}{2023}).
\newblock


\bibitem[Selbst et~al\mbox{.}(2019)]%
        {selbst2019fairness}
\bibfield{author}{\bibinfo{person}{Andrew~D Selbst}, \bibinfo{person}{Danah
  Boyd}, \bibinfo{person}{Sorelle~A Friedler}, \bibinfo{person}{Suresh
  Venkatasubramanian}, {and} \bibinfo{person}{Janet Vertesi}.}
  \bibinfo{year}{2019}\natexlab{}.
\newblock \showarticletitle{Fairness and abstraction in sociotechnical
  systems}. In \bibinfo{booktitle}{\emph{Proceedings of the conference on
  fairness, accountability, and transparency}}. \bibinfo{pages}{59--68}.
\newblock


\bibitem[Sohl-Dickstein et~al\mbox{.}(2015)]%
        {sohldickstein2015deep}
\bibfield{author}{\bibinfo{person}{Jascha Sohl-Dickstein},
  \bibinfo{person}{Eric~A. Weiss}, \bibinfo{person}{Niru Maheswaranathan},
  {and} \bibinfo{person}{Surya Ganguli}.} \bibinfo{year}{2015}\natexlab{}.
\newblock \bibinfo{title}{Deep Unsupervised Learning using Nonequilibrium
  Thermodynamics}.
\newblock
\newblock
\showeprint[arxiv]{1503.03585}~[cs.LG]


\bibitem[Todorov(2019)]%
        {todorov2019game}
\bibfield{author}{\bibinfo{person}{Paul Todorov}.}
  \bibinfo{year}{2019}\natexlab{}.
\newblock \showarticletitle{A game of dice: Machine learning and the question
  concerning art}.
\newblock \bibinfo{journal}{\emph{arXiv preprint arXiv:1904.01957}}
  (\bibinfo{year}{2019}).
\newblock


\bibitem[Tomlinson et~al\mbox{.}(2023)]%
        {tomlinson2023chatgpt}
\bibfield{author}{\bibinfo{person}{Bill Tomlinson}, \bibinfo{person}{Andrew~W
  Torrance}, {and} \bibinfo{person}{Rebecca~W Black}.}
  \bibinfo{year}{2023}\natexlab{}.
\newblock \showarticletitle{ChatGPT and Works Scholarly: Best Practices and
  Legal Pitfalls in Writing with AI}.
\newblock \bibinfo{journal}{\emph{arXiv preprint arXiv:2305.03722}}
  (\bibinfo{year}{2023}).
\newblock


\bibitem[Triangulation(2014)]%
        {triangulation2014use}
\bibfield{author}{\bibinfo{person}{Data~Source Triangulation}.}
  \bibinfo{year}{2014}\natexlab{}.
\newblock \showarticletitle{The use of triangulation in qualitative research}.
  In \bibinfo{booktitle}{\emph{Oncol Nurs Forum}}, Vol.~\bibinfo{volume}{41}.
  \bibinfo{pages}{545--7}.
\newblock


\bibitem[Wach et~al\mbox{.}(2023)]%
        {wach2023dark}
\bibfield{author}{\bibinfo{person}{Krzysztof Wach}, \bibinfo{person}{Cong~Doanh
  Duong}, \bibinfo{person}{Joanna Ejdys}, \bibinfo{person}{R{\=u}ta
  Kazlauskait{\.e}}, \bibinfo{person}{Pawel Korzynski},
  \bibinfo{person}{Grzegorz Mazurek}, \bibinfo{person}{Joanna Paliszkiewicz},
  {and} \bibinfo{person}{Ewa Ziemba}.} \bibinfo{year}{2023}\natexlab{}.
\newblock \showarticletitle{The dark side of generative artificial
  intelligence: A critical analysis of controversies and risks of ChatGPT.}
\newblock \bibinfo{journal}{\emph{Entrepreneurial Business \& Economics
  Review}} \bibinfo{volume}{11}, \bibinfo{number}{2} (\bibinfo{year}{2023}).
\newblock


\bibitem[Walsh et~al\mbox{.}(2022)]%
        {10.1145/3478432.3499157}
\bibfield{author}{\bibinfo{person}{Benjamin Walsh}, \bibinfo{person}{Safinah
  Ali}, \bibinfo{person}{Francisco Castro}, \bibinfo{person}{Kayla Desportes},
  \bibinfo{person}{Daniella DiPaola}, \bibinfo{person}{Irene Lee},
  \bibinfo{person}{William Payne}, \bibinfo{person}{Scott Sieke}, {and}
  \bibinfo{person}{Helen Zhang}.} \bibinfo{year}{2022}\natexlab{}.
\newblock \showarticletitle{Making Art with and about Artificial Intelligence:
  Three Approaches to Teaching AI and AI Ethics to Middle and High School
  Students}. In \bibinfo{booktitle}{\emph{Proceedings of the 53rd ACM Technical
  Symposium on Computer Science Education V. 2}} (Providence, RI, USA)
  \emph{(\bibinfo{series}{SIGCSE 2022})}. \bibinfo{publisher}{Association for
  Computing Machinery}, \bibinfo{address}{New York, NY, USA},
  \bibinfo{pages}{1203}.
\newblock
\showISBNx{9781450390712}
\urldef\tempurl%
\url{https://doi.org/10.1145/3478432.3499157}
\showDOI{\tempurl}


\bibitem[Wan et~al\mbox{.}(2019)]%
        {wan2019towards}
\bibfield{author}{\bibinfo{person}{Chia-Hung Wan}, \bibinfo{person}{Shun-Po
  Chuang}, {and} \bibinfo{person}{Hung-Yi Lee}.}
  \bibinfo{year}{2019}\natexlab{}.
\newblock \showarticletitle{Towards audio to scene image synthesis using
  generative adversarial network}. In \bibinfo{booktitle}{\emph{ICASSP
  2019-2019 IEEE International Conference on Acoustics, Speech and Signal
  Processing (ICASSP)}}. IEEE, \bibinfo{pages}{496--500}.
\newblock


\bibitem[Wang et~al\mbox{.}(2020)]%
        {wang2020imaginator}
\bibfield{author}{\bibinfo{person}{Yaohui Wang}, \bibinfo{person}{Piotr
  Bilinski}, \bibinfo{person}{Francois Bremond}, {and} \bibinfo{person}{Antitza
  Dantcheva}.} \bibinfo{year}{2020}\natexlab{}.
\newblock \showarticletitle{Imaginator: Conditional spatio-temporal gan for
  video generation}. In \bibinfo{booktitle}{\emph{Proceedings of the IEEE/CVF
  Winter Conference on Applications of Computer Vision}}.
  \bibinfo{pages}{1160--1169}.
\newblock


\bibitem[Ward(2017)]%
        {ward2017AI}
\bibfield{author}{\bibinfo{person}{Tom Ward}.} \bibinfo{year}{2017}\natexlab{}.
\newblock \showarticletitle{AI Can Now Produce Better Art Than Humans. Here's
  How.}
\newblock \bibinfo{journal}{\emph{Futurism}} (\bibinfo{year}{2017}).
\newblock
\urldef\tempurl%
\url{https://futurism.com/ai-now-produce-better-art-humans-heres-how}
\showURL{%
\tempurl}


\bibitem[Waters(2023)]%
        {waters2023generative}
\bibfield{author}{\bibinfo{person}{Richard Waters}.}
  \bibinfo{year}{2023}\natexlab{}.
\newblock \bibinfo{booktitle}{\emph{Generative AI: how will the new era of
  machine learning affect you?}}
\newblock
\urldef\tempurl%
\url{https://www.ft.com/content/1e34f334-4e73-4677-9713-99f85eed7ba0}
\showURL{%
\tempurl}


\bibitem[Wu et~al\mbox{.}(2023)]%
        {wu2023art}
\bibfield{author}{\bibinfo{person}{Zhuohao Wu}, \bibinfo{person}{Min Fan},
  \bibinfo{person}{Ritong Tang}, \bibinfo{person}{Danwen Ji}, {and}
  \bibinfo{person}{Mohammad Shidujaman}.} \bibinfo{year}{2023}\natexlab{}.
\newblock \showarticletitle{The Art of Artificial Intelligent Generated Content
  for Mobile Photography}. In \bibinfo{booktitle}{\emph{International
  Conference on Human-Computer Interaction}}. Springer,
  \bibinfo{pages}{438--453}.
\newblock


\bibitem[Yenduri et~al\mbox{.}(2023)]%
        {yenduri2023generative}
\bibfield{author}{\bibinfo{person}{Gokul Yenduri}, \bibinfo{person}{Ramalingam
  M}, \bibinfo{person}{Chemmalar~Selvi G}, \bibinfo{person}{Supriya Y},
  \bibinfo{person}{Gautam Srivastava}, \bibinfo{person}{Praveen Kumar~Reddy
  Maddikunta}, \bibinfo{person}{Deepti~Raj G}, \bibinfo{person}{Rutvij~H
  Jhaveri}, \bibinfo{person}{Prabadevi B}, \bibinfo{person}{Weizheng Wang},
  \bibinfo{person}{Athanasios~V. Vasilakos}, {and}
  \bibinfo{person}{Thippa~Reddy Gadekallu}.} \bibinfo{year}{2023}\natexlab{}.
\newblock \bibinfo{title}{Generative Pre-trained Transformer: A Comprehensive
  Review on Enabling Technologies, Potential Applications, Emerging Challenges,
  and Future Directions}.
\newblock
\newblock
\showeprint[arxiv]{2305.10435}~[cs.CL]


\bibitem[Zarifhonarvar(2023)]%
        {zarifhonarvar2023economics}
\bibfield{author}{\bibinfo{person}{Ali Zarifhonarvar}.}
  \bibinfo{year}{2023}\natexlab{}.
\newblock \showarticletitle{Economics of chatgpt: A labor market view on the
  occupational impact of artificial intelligence}.
\newblock \bibinfo{journal}{\emph{Available at SSRN 4350925}}
  (\bibinfo{year}{2023}).
\newblock


\bibitem[Zhang et~al\mbox{.}(2023)]%
        {zhang2023research}
\bibfield{author}{\bibinfo{person}{Mengqi Zhang}, \bibinfo{person}{Cheng Liu},
  {and} \bibinfo{person}{Zi Wang}.} \bibinfo{year}{2023}\natexlab{}.
\newblock \showarticletitle{Research on Copyright of AI Copycat Works}. In
  \bibinfo{booktitle}{\emph{SHS Web of Conferences}},
  Vol.~\bibinfo{volume}{157}. EDP Sciences, \bibinfo{pages}{04005}.
\newblock


\bibitem[Zhu et~al\mbox{.}(2017)]%
        {Zhu_2017_ICCV}
\bibfield{author}{\bibinfo{person}{Jun-Yan Zhu}, \bibinfo{person}{Taesung
  Park}, \bibinfo{person}{Phillip Isola}, {and} \bibinfo{person}{Alexei~A.
  Efros}.} \bibinfo{year}{2017}\natexlab{}.
\newblock \showarticletitle{Unpaired Image-To-Image Translation Using
  Cycle-Consistent Adversarial Networks}. In
  \bibinfo{booktitle}{\emph{Proceedings of the IEEE International Conference on
  Computer Vision (ICCV)}}.
\newblock


\bibitem[Zhu et~al\mbox{.}(2022)]%
        {zhu2022quantized}
\bibfield{author}{\bibinfo{person}{Ye Zhu}, \bibinfo{person}{Kyle Olszewski},
  \bibinfo{person}{Yu Wu}, \bibinfo{person}{Panos Achlioptas},
  \bibinfo{person}{Menglei Chai}, \bibinfo{person}{Yan Yan}, {and}
  \bibinfo{person}{Sergey Tulyakov}.} \bibinfo{year}{2022}\natexlab{}.
\newblock \showarticletitle{Quantized gan for complex music generation from
  dance videos}. In \bibinfo{booktitle}{\emph{European Conference on Computer
  Vision}}. Springer, \bibinfo{pages}{182--199}.
\newblock


\bibitem[Zylinska(2020a)]%
        {zylinska2020ai_not_creation}
\bibfield{author}{\bibinfo{person}{Joanna Zylinska}.}
  \bibinfo{year}{2020}\natexlab{a}.
\newblock \bibinfo{booktitle}{\emph{AI art: machine visions and warped
  dreams}}.
\newblock \bibinfo{publisher}{Open Humanities Press}, \bibinfo{pages}{76}.
\newblock


\bibitem[Zylinska(2020b)]%
        {zylinska2020ai_teach}
\bibfield{author}{\bibinfo{person}{Joanna Zylinska}.}
  \bibinfo{year}{2020}\natexlab{b}.
\newblock \bibinfo{booktitle}{\emph{AI art: machine visions and warped
  dreams}}.
\newblock \bibinfo{publisher}{Open Humanities Press}, \bibinfo{pages}{29}.
\newblock


\end{thebibliography}

\appendix

\end{document}